\documentclass[article]{aa}  
\usepackage{amsmath}
\usepackage{ amssymb}
\usepackage{graphicx}
\usepackage{graphics}
\usepackage{subfigure}
\usepackage{color}
\usepackage{float}
\usepackage{txfonts}
\usepackage{natbib}
\bibpunct{(}{)}{;}{a}{}{,}

\def\teff{\ifmmode T_{\rm eff} \else $T_{\mathrm{eff}}$\fi}

\def\ltsima{$\buildrel<\over\sim$}
\def\lsim{\lower.5ex\hbox{\ltsima}}
\def\nh{\ifmmode N_{\rm HI} \else $N_{\rm HI}$\fi}

\newcommand{\hi}{H{\sc i}}
\newcommand{\hii}{H{\sc ii}}
\newcommand{\ha}{\ifmmode {\rm H}\alpha \else H$\alpha$\fi}
\newcommand{\hb}{\ifmmode {\rm H}\beta \else H$\beta$\fi}
\newcommand{\lya}{\ifmmode {\rm Ly}\alpha \else Ly$\alpha$\fi}

\newcommand{\vexp}{\ifmmode v_{\rm exp} \else $v_{\rm exp}$\fi}
\newcommand{\vsep}{\ifmmode V_{\rm sep} \else $V_{\rm sep}$\fi}
\newcommand{\vpeak}{\ifmmode v_{\rm peak} \else $v_{\rm peak}$\fi}

\def\fesclyc{$f_{\mathrm{esc}}^{\mathrm{LyC}}$}
\def\fesclya{$f_{\mathrm{esc}}^{\mathrm{\lya}}$}

\def\msun{$M_\odot$}

\def\kms{km s$^{-1}$}

\def\hii{\ion{H}{ii}}
\def\oii{[\ion{O}{ii}]}

\def\oiii{[\ion{O}{iii}]}
\def\oiiil{[\ion{O}{iii}]$\lambda 5007$}

\begin{document}

\title{Lyman-$\alpha$ spectral properties of five newly discovered \\ 
Lyman continuum emitters}

\author{A. Verhamme\inst{1}, 
I. Orlitov\'a\inst{2}, 
D. Schaerer\inst{1,3}, 
Y. Izotov\inst{4}, 
G. Worseck\inst{5}, 
T. X. Thuan\inst{6},
N. Guseva\inst{4}
}
\offprints{anne.verhamme@unige.ch}
\institute{
Observatoire de Gen\`eve, Universit\'e de Gen\`eve, 
51 Ch. des Maillettes, 1290 Versoix, Switzerland
\and
Astronomical Institute, Czech Academy of Sciences,
Bo\v cn{\'\i} II 1401, 141 00 Prague, Czech Republic
\and
CNRS, IRAP, 14 Avenue E. Belin, 31400 Toulouse, France
\and
Main Astronomical Observatory, Ukrainian National Academy of Sciences,
27 Zabolotnoho str., Kyiv 03680, Ukraine
	\and
Max-Planck-Institut f\"ur Astronomie, K\"onigstuhl 17, 69117 Heidelberg, Germany		
	\and
Astronomy Department, University of Virginia, P.O. Box 400325, 
Charlottesville, VA 22904-4325, USA
}
\date{Received date / Accepted date}
\authorrunning{Verhamme et al.}{ }
\titlerunning{\lya\ as a proxy for LyC}{ }

\abstract{}
{We have recently reported the discovery of five low redshift Lyman continuum (LyC) emitters (LCEs, hereafter)
with absolute escape fractions \fesclyc\ ranging from 6 to 13\%, higher than previously found, and which more than doubles the number of low redshift LCEs.
We use these observations to test theoretical predictions about a link between the characteristics of the Lyman-alpha (\lya) line from galaxies and the escape of ionising photons.
} 
{We analyse the \lya\ spectra of eight LCEs of the local Universe observed with the Cosmic Origins Spectrograph onboard the Hubble Space Telescope
(our five leakers and three galaxies from the litterature), 
and compare their strengths and shapes to the theoretical criteria and comparison samples of local galaxies: the Lyman Alpha Reference Survey, Lyman Break Analogs, Green Peas, and the high-redshift strong LyC leaker {\it Ion2}.}
{Our LCEs are found to be strong \lya\ emitters, with high equivalent widths, EW(\lya)$> 70$ \AA, and large \lya\ escape fractions, \fesclya $>20\%$. The \lya\ profiles are all double-peaked with a small peak separation, in agreement with our theoretical expectations. They also have no underlying absorption at the \lya\ position. All these characteristics are very different from the \lya\ properties of typical star-forming galaxies of the local Universe. A subset of the comparison samples (2-3 Green Pea galaxies) share these extreme values, indicating that they could also be leaking. We also find a strong correlation between the star formation rate surface density and the escape fraction of ionising photons, indicating that the compactness of star-forming regions plays a role in shaping low column density paths in the interstellar medium of LCEs.
} 
{ The \lya\ properties of LCEs are peculiar: \lya\ can be used as a reliable tracer of LyC escape from galaxies, in complement to other indirect diagnostics proposed in the literature.}

\keywords{Radiative transfer -- dark ages, reionization, first stars -- Galaxies: ISM -- ISM: structure -- 
ISM: kinematics and dynamics}

\maketitle

\section{Introduction}
\label{s_intro}

Cosmic reionisation is a major event in the history of the Universe, which took place during the first billion years: after the dark ages, the first sources of light released ionising radiation in the intergalactic medium, reheating the gas, and changing its opacity. This \emph{cosmic feedback} of the first astrophysical objects is thought to have influenced the formation and evolution of galaxies in their first stages \citep{Ocvirk16} and determines their detectability \citep{Stark11, Schenker14, Dijkstra14, Dijkstra15}. The open question in actual studies of cosmic reionisation is to understand the relative contribution of the two types of sources likely involved: quasars and star-forming galaxies. Quasars inject a copious amount of ionising photons into the intergalactic medium (IGM), but were thought to be too rare at high redshift to have contributed significantly \citep{Fontanot12,Fontanot14}. A new population of faint Active Galactive Nuclei (AGNs) has been recently reported \citep{Giallongo15}, which could sustain reionisation, assuming that their escape fraction of ionising photons is high \citep[e.g. 100\%\ in ][]{Madau15}. The escape fraction of ionising photons from AGNs has to be investigated in more details, recent measurements report much lower values than assumed in the previous studies \citep{Micheva16}. 

Our present study focuses on the second type of sources of the cosmic reionisation: 
massive stars in galaxies. As for AGNs, the main uncertainty to quantify the role of star formation in the cosmic reionisation is the escape fraction of ionising photons from galaxies. Young and massive stars produce ionising radiation in situ, the difficult question is to observe, measure, and quantify how many of these photons, if any, escape the interstellar medium (ISM) of galaxies. So far, one clear detection has been reported at high redshift, with a high Lyman continuum absolute escape fraction \citep[\fesclyc $>50$\%][]{Vanzella16, deBarros16}, another  ``clean'' Lyman Continuum Emitter (LCE) was recently found at $z=3.15$ by \cite{Shapley16}, and few low-redshift detections with low escape fractions ($<5\%$) have been found during the last decade \citep[][Puschnig et al., submitted]{Bergvall06,Leitet13,Borthakur14,Leitherer16}. 
Significant progress at low redshifts has recently been achieved with the COS spectrograph onboard {\sl HST} by 
\cite{Izotov16a, Izotov16b}, who found five LCEs at $z \sim 0.3$ with \fesclyc $\sim 6-13$\%.  
These sources are the basis of the present study.

Given the extremely low success rate of observational searches for LCEs, several pre-selection methods for good LCE candidates have been proposed,
involving the alteration of nebular emission line strengths \citep{Zackrisson13}, high \oiii/\oii\ ratios potentially tracing density-bounded \hii\ regions \citep{Jaskot13, Nakajima14}, or the non-saturation of the metallic low-ionisation absorption lines \citep{Heckman11, Alexandroff15, Vasei16} tracing a low covering fraction of the absorbing gas along the line of sight. We recently proposed a method based on the Lyman-alpha (\lya) spectral shape of LCEs: strong and narrow \lya\ profiles with a shift of the main peak smaller than $\sim150$ \kms, or double peaks closer than $\sim300$ \kms would indicate LyC leakage \citep[see ][ for more details]{Verhamme15}. Time is ripe to test these simple predictions. 

In the present work we assess the specificity of the \lya\ properties of LCEs by comparison to other samples of low-redshift star-forming galaxies whose \lya\ properties have been measured and analysed:
the Lyman Alpha Reference Survey (LARS), Lyman-Break Analogs (LBAs), and ``Green Pea'' galaxies (GPs). 
The LARS sample consists of 14 star-forming galaxies at redshifts $z = 0.02 - 0.2$.
They were selected from the cross-matched SDSS and {\sl GALEX} catalogues to sample the range of far-UV luminosities observed in $z\sim3$ Lyman Break Galaxies (LBGs). Only galaxies with active star formation were included, requiring EW(Ha)$ > 100$\,\AA. The aim of LARS is to study the mechanisms governing \lya\ escape from galaxies \citep[see][for an overview]{Ostlin14}. Most of LARS galaxies are dwarf irregulars. LARS14 is also a GP (see below).  

The LBAs were selected from the {\sl GALEX} catalogue for their far-UV luminosity, high surface brightness, and compactness \citep{Heckman05}. They resemble the high-redshift LBGs in physical size, stellar mass, star-formation rate, metallicity, dust extinction, and gas velocity dispersion. UV and optical morphologies of 30 LBAs were studied by \citet{Overzier09}.
The UV shows massive star-forming clumps, while evidence of interaction is seen in the optical images. In 20\% of the sample, all of the UV light comes from a single, compact star-forming clump, usually characterized by strong outflows. Far-UV spectra of 22 LBAs were obtained with {\sl HST}/COS \citep{Heckman11, Alexandroff15}.
Most of their \lya\ lines are observed in emission, with a variety of profiles including P-Cygni, as well as broad and narrow double-peaks. The escape of ionising photons from one LBA, J0921+4509, was reported in \citet{Borthakur14}, with an absolute escape fraction of $1\%$. The probability of LyC leakage from the other LBAs was discussed in \citet{Alexandroff15} by comparing the indirect diagnostics for LyC escape discussed above.

Green Pea galaxies were noticed in the Galaxy Zoo SDSS images for their bright green colour and compactness. The colour is due to strong \oiiil\,\AA\ emission, reaching equivalent widths as high as 1000\,\AA. \citet{Cardamone09} identified a sample of $\sim250$ GPs in SDSS DR7, at $z=0.11-0.36$, while \citet{Izotov11} extended the number to $\sim800$ over a larger redshift range $z=0.02-0.63$ (the colour changes with redshift, but their properties remain similar). GPs share many properties with high-redshift LBGs and LAEs. Twelve archival {\sl HST}/COS spectra of GPs (outside our sample) are available \citep{Henry15,Yang16}, drawn from \citet{Cardamone09}.

They are all strong \lya\ emitters, most of them with double-peaked line profiles. The escape of ionising photons from GPs was discussed on the basis of their \lya\ spectral shapes \citep{Verhamme15}. Some are best fitted by synthetic \lya\ spectra emergent from low column density geometries, indicating that they could be leaking \citep[][Orlitova et al. in prep.]{Yang16}.

As reported in \citet{Izotov16a, Izotov16b}, we detected LyC emission from five compact, strongly star-forming galaxies of the local Universe, with high \oiii/\oii\ ratios ($>4$). These galaxies belong to the GP category. To our knowledge, this was the first attempt to test if high \oiii/\oii\ ratios are linked to LyC leakage. The major result of this study has been that five out of the five observed objects are leaking ionising radiation, and the finding of a correlation between \fesclyc\ and \oiii/\oii\ \cite[see Fig.\ 14 in][]{Izotov16b}. Therefore, a high \oiii/\oii\ ratio appears as a potential signature of LyC escape. Since nothing was known about \lya\  for these galaxies, it is interesting to study the \lya\ properties of these LCEs, and to use these sources to test our theoretical predictions regarding the relation between LyC leakage, \lya\ escape, and the detailed \lya\ line profiles \citep{Verhamme15}. This is the main objective of this paper.

In Sect.~\ref{s_data} we describe the observational data used in our study.
In Sect.~\ref{s_LyaLyC} we discuss the \lya\ properties of LCEs and comparison samples, showing that LCEs have a strong \lya\ emission.
We then present and discuss the detailed \lya\ line profiles of these galaxies (Sect.~\ref{s_profiles}).
The results are briefly put into perspective and discussed in Sect.~\ref{s_discuss}. 
Our main conclusions are summarised in Sect.~\ref{s_conclude}.

\section{Description of the data}
\label{s_data}

\begin{table*}
  \caption{Measured \lya\ and other related properties of all known low-$z$ LCEs, ordered by decreasing \fesclyc (8th column). The \lya\ equivalent width, EW(\lya), is given in the rest-frame. The EW uncertainties are estimated as $\pm 20$\AA. The last column contains the star formation rate per unit area, $\Sigma_{\mathrm{SFR}}$, in \msun\,yr$^{-1}$\,kpc$^{-2}$. Columns 2, and 8-10 are taken from  \cite{Izotov16a, Izotov16b}. Note that \fesclya\ has been recalculated from \citet{Izotov16a,Izotov16b}, without SDSS vs COS aperture correction, for comparison with other samples (see text for more details). The typical uncertainties on velocity shift measurements are $\pm 10$ \kms (see text for more details).
}
\begin{tabular}{cccccccccc}
\hline
ID &  EW(\lya)  & \fesclya\ & $V_{\mathrm{peak}}^{\mathrm{blue}}$ & $V_{\mathrm{peak}}^{\mathrm{red}}$ & \vsep\ & EW$_{\mathrm{blue}}/$EW$_{\mathrm{red}}$ & \fesclyc & \oiii/\oii & $\Sigma_{\mathrm{SFR}}$ \\
  &  \AA & & \kms & \kms & \kms &  & & & \msun\,yr$^{-1}$\,kpc$^{-2}$ \\
\hline
\object{J1152+3400}  &   79 & 0.36 & -120 & 150 & 270 & 0.55 & 0.132 & 5.4 & 35.5 \\
\object{J1442$-$0209}&  129 & 0.57 & -250 &  60 & 310 & 0.17 & 0.074 & 6.7 & 15.5 \\
\object{J0925+1403}  &   83 & 0.31 & -160 & 150 & 310 & 0.40 & 0.072 & 4.8 & 12.6 \\
\object{J1503+3644}  &   98 & 0.31 & -290 & 140 & 430 & 0.17 & 0.058 & 4.9 &  6.8 \\
\object{J1333+6246}  &   75 & 0.55 & -300 &  90 & 390 & 0.10 & 0.056 & 4.8 &  2.2 \\
\hline
\object{Tol 1247-232}&   29 & 0.20 & -300 & 150 & 450 & 0.12 & 0.045 & 3.4 &      \\
\object{Haro11}      &   15 & 0.04 & -300 & 110 & 410 & 0.06 & 0.032 & 1.5 &  1.1 \\
\object{J0921+4509}  &    4 & 0.01 & -450 & 240 & 690 & 1.0  &  0.01 & 0.3 &  7.9 \\
\hline
\end{tabular}
\label{table_LyaLyC}
\end{table*}

\subsection{Lyman continuum leakers}
We use \lya\ observations of all known low redshift LCEs combining data from five objects recently reported
by our team  \citep{Izotov16a, Izotov16b} and three other sources from the literature.
The \lya\ spectra of our LCEs were observed with {\sl HST}/COS using two gratings: the medium resolution G160M grating, and the low resolution G140L grating (GO 13744, PI: T.X.Thuan). The data were reduced with a custom software specifically designed for faint {\sl HST}/COS targets \citep{Worseck11, Syphers12}, as described in  \cite{Izotov16a, Izotov16b}. 
For \ \lya\ \ and other parts of the spectrum with strong fluxes our custom reduction yields results which are equivalent to those of the standard COS pipeline.
To this we add the other three low redshift objects which have a direct detection in the Lyman continuum and whose \lya\ spectra have also been observed with COS. Haro 11 is a known LyC leaker of the local Universe, with an absolute escape fraction \fesclyc$\sim 3\%$ \citep{Leitet11, Leitet13}, its \lya\ spectrum has been observed with the G130M grating (GO13017, PI: T.Heckman) and discussed in  \cite{Alexandroff15} and \cite{Verhamme15}. 
Tololo 1247-323 has \fesclyc$\sim 4-6\%$ \citep[][Puschnig et al submitted]{Leitet13, Leitherer16}, and its \lya\ spectrum was observed with G130M (GO13027, PI: G.Ostlin, Puschnig et al submitted). Ionising flux has also been detected from the LBA J0921+4509, with \fesclyc$\sim1\%$ \citep{Borthakur14}, and its \lya\ spectrum has also been observed with G130M \citep[GO11727, PI: T.Heckman,][]{Heckman11, Verhamme15, Alexandroff15}.
The COS observations of the three sources from the literature were retrieved from the MAST archive, reduced with the up-to-date version of the standard COS pipeline.

All spectra were binned in the same way, and the following quantities were measured in a homogeneous manner: the \lya\ equivalent width, EW(\lya), the location of local maxima in the profiles, $V_{\mathrm{peak}}^{\mathrm{blue}}$ for the maximum blueward of the systemic redshift, and $V_{\mathrm{peak}}^{\mathrm{red}}$ for the maximum on the red side. We also measure the separation between the red and blue peaks: 
\begin{equation} 
\vsep  = V_{\mathrm{peak}}^{\mathrm{red}} - V_{\mathrm{peak}}^{\mathrm{blue}}.
\end{equation}
The measurements for all LCEs are summarised in Table~\ref{table_LyaLyC}.
The effective spectral resolution of the spectra is $R \sim 4000$, as estimated by convolving the observed spectrum with gaussians to mimic the effect of the line spread function on the data.  

These measurements are affected by spectral resolution, as will be demonstrated further, but maybe also by the sampling of the data, the continuum determination, and the redshift accuracy. 
Since our galaxies are in the SDSS database, the data for several strong optical nebular lines are available, allowing for a very precise systemic redshift measurement ($<15$ \kms, for a typical spectral resolution of R$\sim 2000$ in SDSS). The main uncertainty on velocity shift measurements comes from COS wavelength calibration, which we estimate at $\pm 40$ \kms, following the detailed discussion in \citet{Henry15}. 
The \lya\ sizes of galaxies are usually bigger than their UV size \citep{Steidel11, Hayes13}. COS resolution decreases with the extent of the source inside the aperture, so the effective spectral resolution around \lya\ is probably lower than in the continuum. We binned the data to increase the signal to noise, keeping $\sim5$ resolution elements in the main component of the profile, i.e. the red peak. We also checked that different binnings lead to velocity shifts much lower than the wavelength calibration uncertainties, $\Delta V \sim \pm 10$ \kms.
We measured the continuum level from two different methods (local estimation vs. global fit). The difference in EW measurements computed from these two continuum estimates is typically $\pm 20$ \AA, which we assume to be the typical uncertainty on our EW measurements. 

The \lya\ escape fractions, reported in Table \ref{table_LyaLyC}, were calculated from the Balmer lines in the SDSS spectra for our five LCEs by \cite{Izotov16b} and J0921+04509, by the equation:
\begin{equation}
$\fesclya$ =  F(\lya)/  ( 8.2 \times F(H\alpha)_{\rm corr} ) , 
\end{equation} 
where F(\lya) is the observed \lya\ flux, and F(H$\alpha)_{\rm corr}$ the H$\alpha$ flux corrected for internal extinction. Both \lya\ and H$\alpha$ are corrected for the foreground Milky Way extinction. We assumed an intrinsic factor of 8.2 between the intrinsic\lya\ and H$\alpha$ fluxes \citep{Dopita03}. We could use the H$\alpha$ SDSS data because the SDSS aperture is of similar size of the {\sl HST}/COS spectrograph (2.5" for COS and 3" for SDSS). In contrast with the measures reported in \citet{Izotov16a,Izotov16b}, we do not apply any aperture correction here. We compute \fesclya\ from our LCEs, in the same way as in other studies, in order to compare with values reported in the literature, GPs in particular. Therefore our \fesclya\ values are larger by a factor 1.4 compared to those reported in \citet{Izotov16a,Izotov16b}. 
The two other LCEs Haro 11 and Tol 1247-232 are not in the SDSS database, so \fesclya\ are derived from imaging and not spectroscopy, they are integrated \fesclya\ measurements (as for LARS) and not limited to the COS aperture. For Haro 11 \cite{Ostlin09} have obtained \fesclya $= 0.037$ from \lya\ and H$\alpha$ images. A value of \fesclya $=0.20$ has been derived by Puschnig et al.\ (private communication) for Tol 1247-232 from {\sl HST} imaging and COS spectroscopy.

\subsection{Comparison samples}

Subsequently we will compare our measurements with the \lya\ properties of two other samples of local galaxies, which have been discussed in the context of LyC leakage from galaxies, the GPs and LBAs. In practice we have been able to compile measurements for 12 GPs \citep{Jaskot14,Henry15,Yang16}, and 8 LBAs \citep{Heckman11}. The \lya\ measurements are taken from these papers and from \citet{Verhamme15}. For comparison with ``normal'' low redshift star-forming galaxies we also use \lya\ data of the 14 LARS galaxies reported in \citet{Hayes13, Hayes14, Rivera15}. In particular, the LARS EW(\lya) and \fesclya\ reported in Fig. 1 are global measurements, derived from \lya\ and H$\alpha$ imaging, taken from \citep{Hayes14}. The LARS, LBA and GP samples have briefly been discussed above (Sect.~\ref{s_intro}).
Finally, we also include the confirmed $z=3.218$ Lyman continuum leaker {\it Ion2} from \cite{Vanzella15}, \cite{deBarros16}, and \cite{Vanzella16} for comparison. From these papers, {\it Ion2} has EW(\lya)$\sim 94$\AA, \fesclyc $\sim 0.64$, \fesclya $>0.78$, \oiii/\oii$>10$ and a SFR density $\Sigma_{\mathrm{SFR}} \sim 43$ \msun.yr$^{-1}$.kpc$^{-2}$.

\begin{figure}[tb]
\includegraphics[width=0.5\textwidth]{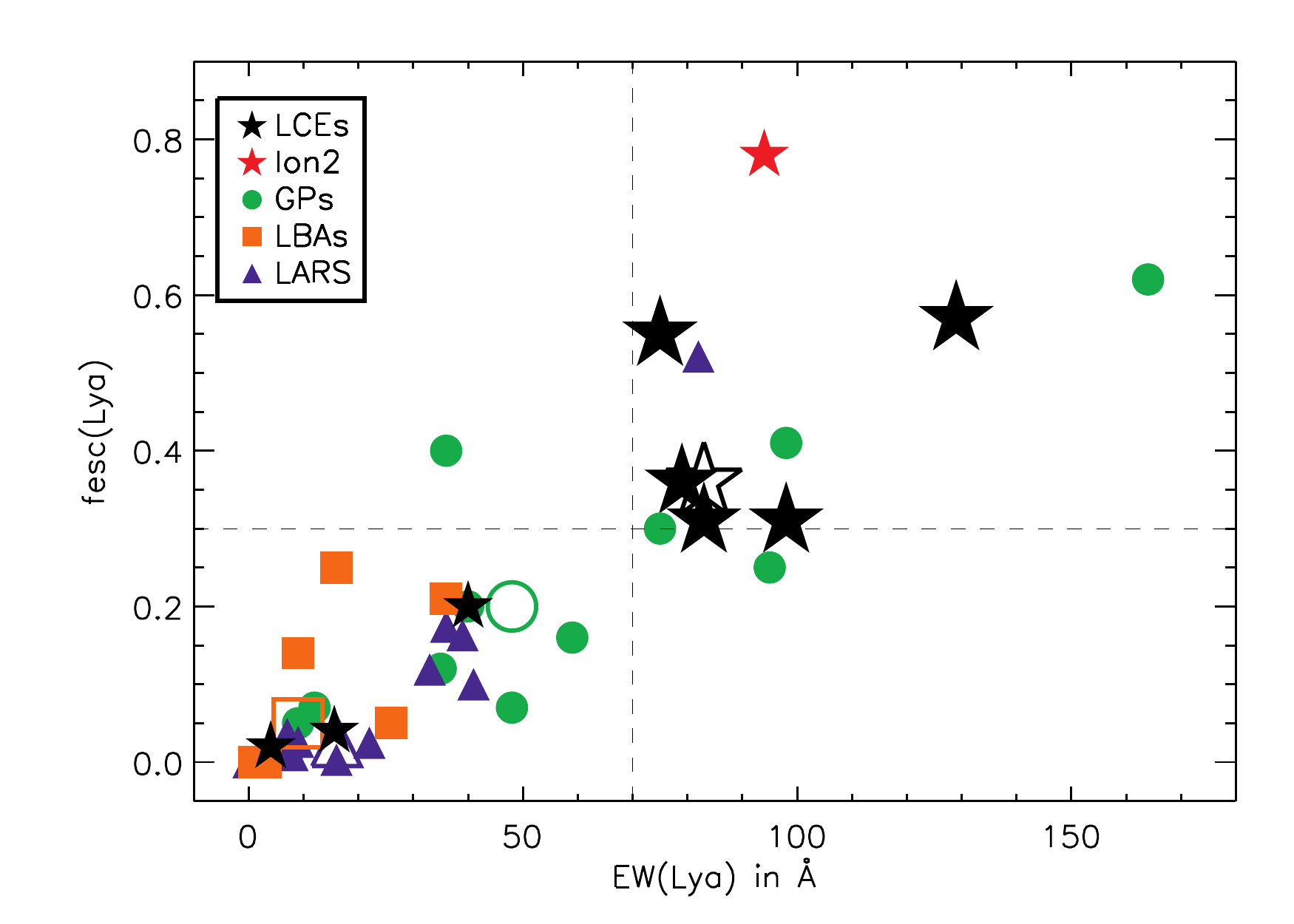} \\
\caption{Comparison of two measures of the \lya\ strength, \fesclya\ and EW(\lya), of LCEs with that of GPs, LBAs and LARS: strong LCEs, shown by large black stars, are more extreme than the comparison samples. The open symbols show the median of the distributions, for each sample. 
The red star shows the location of the high-redshift strong LyC leaker {\it Ion2}. 
The dashed lines delimit the upper right panel containing all strong LyC leakers.}
\label{LyaStrength}
\end{figure}

\section{LCEs have strong \lya\ emission}
\label{s_LyaLyC}

In Fig.~\ref{LyaStrength} we plot the \lya\ equivalent width, EW(\lya), of the eight low-redshift LCEs from Table \ref{table_LyaLyC} as well as the high-$z$ LCE {\it Ion2} (star symbols) as a function of \fesclya. The comparison samples, including GPs, LBAs, and the LARS sources are also shown.
Clearly, the five compact LCEs from \cite{Izotov16a} and \cite{Izotov16b} fall in the upper right corner of the plot: they have \fesclya$>30\%$ and EW(\lya)$>70$\AA. These are extreme values for galaxies of the local Universe, very different from the distribution of EW(\lya) for LARS and LBAs, whose median values are shown by the large open symbols. Only some GPs (and one LARS object, LARS02) fall into the same region of the plot, and their median value is higher. {\it Ion2} (red star), the strong LyC leaker of the high-z Universe \citep{deBarros16, Vanzella16}, also falls into the same region. 

As shown by the open black star, the median values of our 5 LCEs are \fesclya$ = 0.36$ and EW(\lya)$ = 83$\AA, for a median \fesclyc$ = 0.07$. So empirically, we deduce from Fig.~\ref{LyaStrength} that LCEs with \fesclyc $>5\%$ have a high \fesclya ($>0.30$) and a high EW ($>70$\AA). However, whether the reverse is also true, i.e.\ if all galaxies with high \fesclya\ and high EW(\lya) are also leaking ionising photons, remains to be examined. 
Theoretically, all dust-free systems have an angle-averaged \fesclya $=1$, since \lya\ photons are only scattered and not destroyed by interactions with hydrogen, independently of their LyC optical depth. 
Galaxies with a high observed \lya\ equivalent width but a low \fesclya\ (i.e.\ in the lower right corner of  Fig.~\ref{LyaStrength}), are not expected, as this would imply a very large intrinsic EW(\lya), whereas a maximum EW(\lya) $\approx$ 240 \AA\ is expected for very young, normal stellar populations \citep{Schaerer03}.

\begin{figure*}[htb]
\begin{tabular}{cc}
\includegraphics[width=0.45\textwidth]{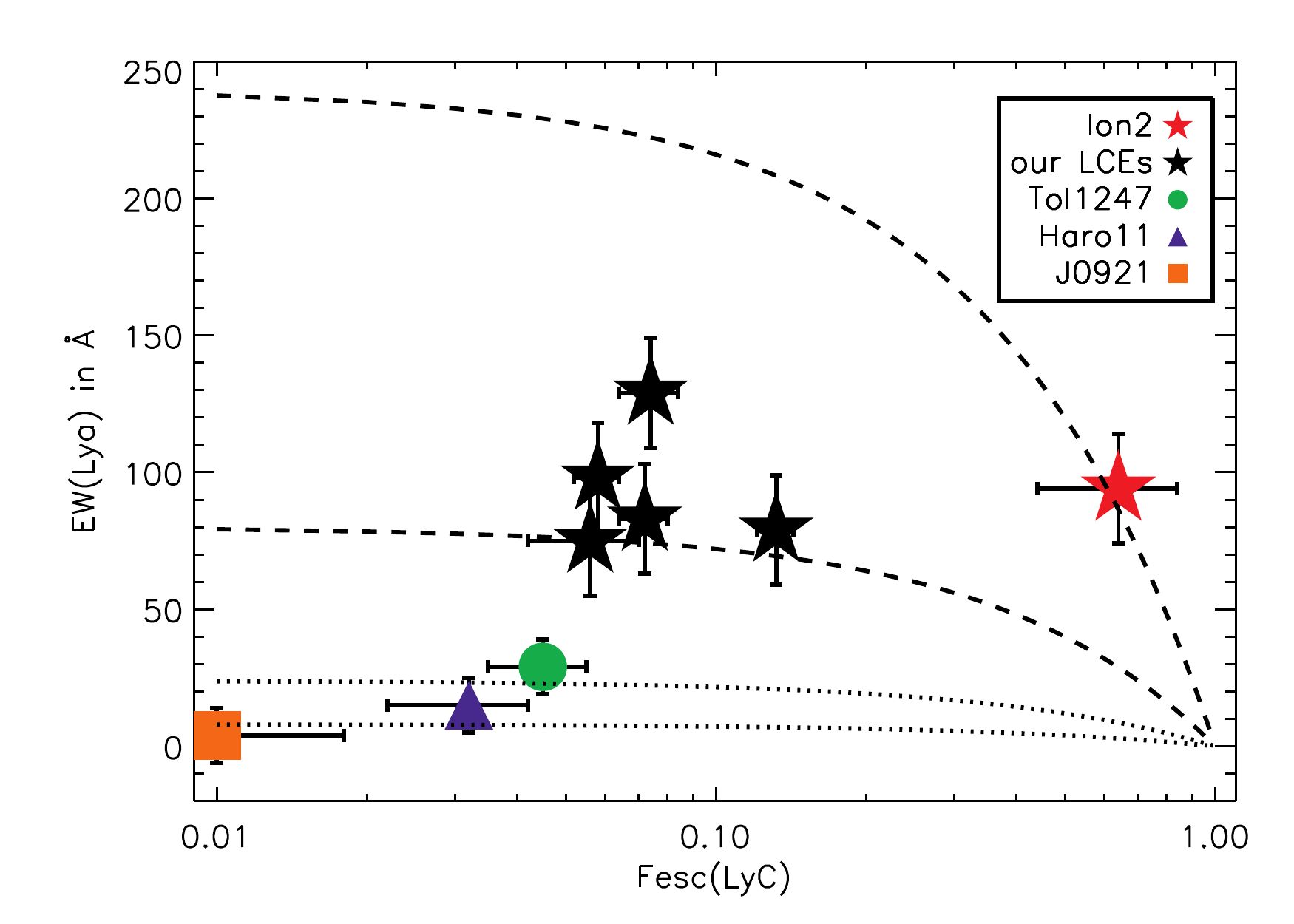} &
\includegraphics[width=0.45\textwidth]{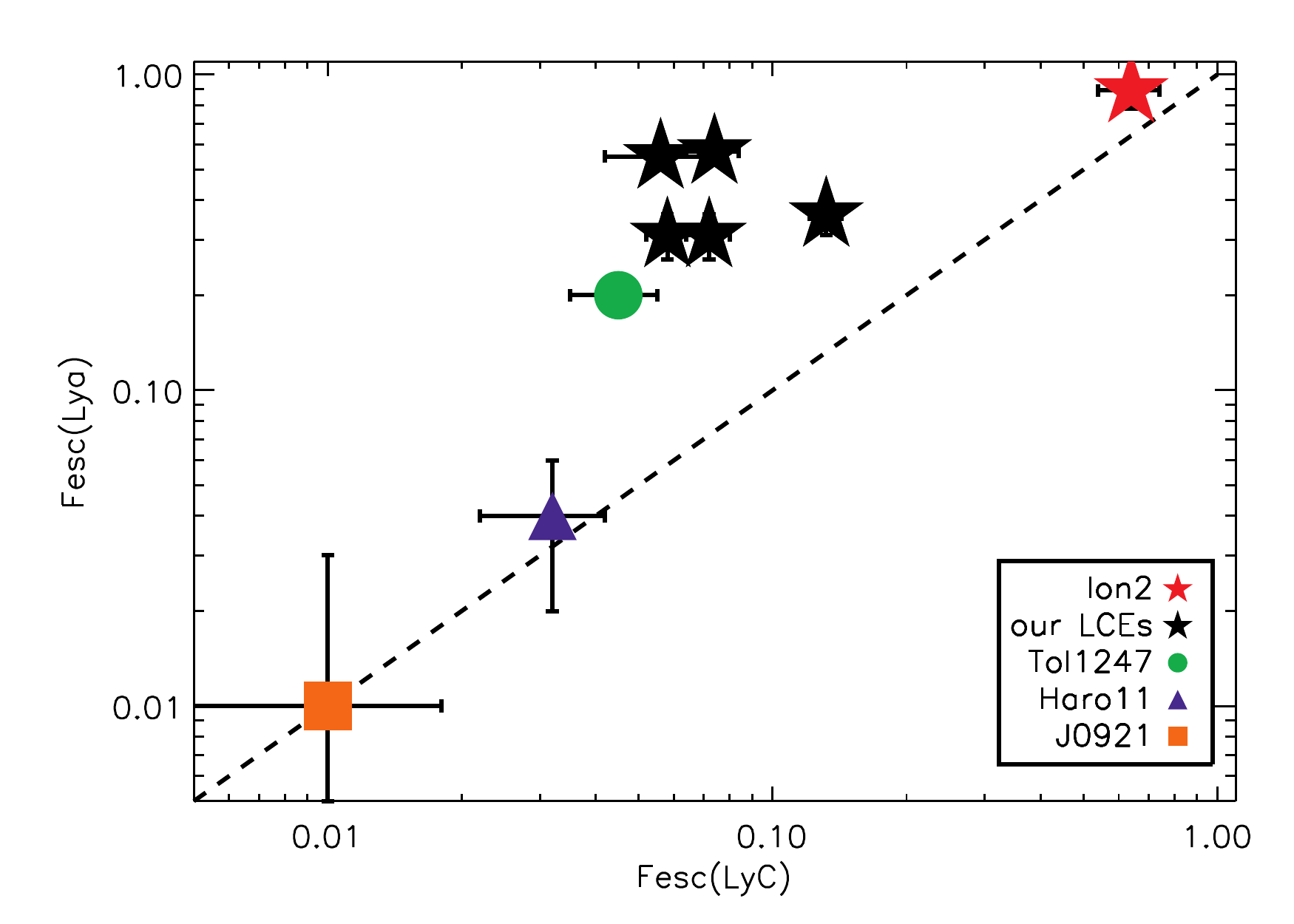} \\
\end{tabular}
\caption{\lya\ strength versus escape of ionising photons from eight galaxies of the local Universe: our five leakers \citep{Izotov16a, Izotov16b}, the LBA J0921+4509 \citep{Borthakur14}, Haro11 and Tol1247\citep{Leitherer16}, and the high redshift galaxy {\it Ion2} \citep{deBarros16, Vanzella16}. {\bf Left:} The \lya\ EWs of LCEs seem to correlate with the escape fraction of ionising photons. The dashed lines show theoretical predictions for the same attenuation in the \lya\ line than in the continuum, for an instantaneous burst (upper curve) or a constant star formation history (lower curve). The dotted lines are the same, but for \lya\ ten times more extinguished than the continuum.
See text for more details.  
{\bf Right:} \fesclya correlates with \fesclyc, and is generally larger than \fesclyc. The dashed line shows the one to one relation. 
}
\label{LyaStrength_vs_LyC}
\end{figure*}

We now investigate direct correlations between \fesclyc\ and the two indicators of strength of the \lya\ emission, EW(\lya) and \fesclya, for the LCE sample.
On the left panel of  Fig.~\ref{LyaStrength_vs_LyC}, we find a trend between EW(\lya) and \fesclyc : galaxies with higher \fesclyc have higher EW(Lya). 
Part of this observed trend could be due to selection effects, as the LCEs from our team (black stars) were selected for having strong emission lines in the optical (among other criteria) to search for sources with an intrinsically strong Lyman continuum flux. In principle sources with faint emission lines, e.g.\ at the advanced stage after a burst of star formation, could have an ISM allowing the escape of LyC photons, which would correspond to sources with a low EW(\lya) and arbitrary values of \fesclyc. Whether such sources exist in nature can presently not be excluded. However,  their contribution to cosmic reionisation would probably be insignificant, since they would have a low LyC photon production. In other words, it seems more likely that sources of relevance for cosmic reionisation, say with \fesclyc $\ga$ 5\%, show also a high \lya\ equivalent width, as the sources shown in  Fig.~\ref{LyaStrength_vs_LyC}.

For a given intrinsic \lya\ equivalent width of the population, EW$_{\mathrm{int}}$(\lya), the observed EW(\lya) and \fesclyc\ are linked by the equation: 
\begin{equation}
EW(\lya) = EW_{\mathrm{int}}(\lya) \times (1 - f_{\mathrm{esc}}^{\mathrm{LyC}}) \times f_{\mathrm{esc}}^{\mathrm{\lya}} / f_{\mathrm{esc}}^{\mathrm{UV}},
\label{eq_ew}
\end{equation}
where $f_{\mathrm{esc}}^{\mathrm{UV}}$ is the escape fraction of the non-ionising, non-resonant UV continuum around \lya, and \fesclya\ is as before. 
For two maximum values of intrinsic \lya\ emission limits with EW$_{\mathrm{int}}$(\lya)=240 and 80 \AA\ respectively, corresponding to a young metal-poor burst and constant star-formation \citep[cf.][]{Schaerer03}, the dashed lines in  Fig.~\ref{LyaStrength_vs_LyC} show the corresponding maximum observed EW(\lya) assuming $f_{\mathrm{esc}}^{\mathrm{\lya}} = f_{\mathrm{esc}}^{\mathrm{UV}}$. This assumption implies that \lya\ photons are destroyed by dust in the same proportion as the UV non-resonant radiation, as observed for at least some local LAEs \citep{Atek14, Henry15}. This shows that observed EWs of the strong LCEs are within the expected range. Only for very high values of the Lyman continuum escape (\fesclyc $\ga 0.5$) does the expected \lya\ strength also decrease due to a significant loss of ionising photons, as illustrated by the decrease of the dashed lines in Fig.~\ref{LyaStrength_vs_LyC} \citep[ see also Fig.13 in][]{Nakajima14}.
The dotted lines in  Fig.~\ref{LyaStrength_vs_LyC} show the same, but for $f_{\mathrm{esc}}^{\mathrm{\lya}} = 0.1 f_{\mathrm{esc}}^{\mathrm{UV}}$, i.e.\ when \lya\ photons are more destroyed by dust than UV continuum photons. This effect, observed in many star-forming galaxies \citep[e.g.][]{Atek08,Atek14}, is one possible explanation for the low EW(\lya) observed in three of the low-$z$ LCEs.

On the right panel of Fig~\ref{LyaStrength_vs_LyC}, we show  \fesclya\ as a function of \fesclyc, which are found to correlate for the eight LCEs and {\it Ion2}. As the dashed line indicating the one-to-one relation shows, the \lya\ escape fraction is generally higher than the Lyman continuum escape fraction, typically $\sim 2-10$ times higher for this set of data, except for Haro11. This result can be explained as follows: the escape of ionising photons from galaxies is only possible when the amount of neutral gas along the line of sight is particularly low ($\log(\nh <1.6 \times 10^{17}$ cm$^{-2}$), at least over a partial area of the galaxy in front of star-forming regions. Furthermore, this peculiar geometrical configuration is also extremely favorable to \lya\ escape \citep{Behrens14, Verhamme15}, so that \lya\ photons can preferentially escape along these low column density channels. The reason why \fesclya\ is expected to be higher than \fesclyc\ is due to the resonant nature of the line: thanks to scattering, some \lya\ photons which were not emitted in front of a hole can be scattered into a clear line of sight, so that more \lya\ photons will find holes compared to the non-resonant LyC photons. Therefore, LCEs are expected to have a high \fesclya\ and \fesclya$/$ \fesclyc$>1$. If our empirical result holds more universally, it implies that objects with \fesclyc $>10\%$ should have at least \fesclya\ $>20\%$. Larger samples will be needed to firm up this result.

Regarding the observed relation between \fesclya\ and \fesclyc\ we also note that theoretical predictions from clumpy geometries presented by \citet{Dijkstra16} show the same trend of \fesclya $>$ \fesclyc, albeit with a much larger scatter in the ratio \fesclya$/$\fesclyc\ than observed here. However, their predicted escape fractions correspond to averages over all angles, i.e.\  ``global'' escape fractions, which are not directly comparable to observations.
Finally, we have searched the literature for other samples providing measurements (or upper limits) of both \fesclya\ and \fesclyc. The few studies/surveys providing this data \citep[e.g.\ $z \sim 2$ H$\alpha$ emitters,][]{Matthee16a, Matthee16b} are compatible with the relation between \fesclya\ and \fesclyc\ followed by the LCEs studied here. In particular, no strong LCE with a low \lya\ escape fraction is known, nor do we know of sources with a high \fesclya\ and a stringent LyC detection limit.

We conclude that the current data shown in both panels of  Fig.~\ref{LyaStrength_vs_LyC} strongly suggest that strong LCEs should be found among galaxy samples with strong \lya\ emission, characterised by a high equivalent width (EW(\lya) $\ga 70$ \AA), and/or a high \lya\ escape fraction. We now examine the detailed \lya\ line profiles of the LCEs.

\begin{figure}[htb]
\includegraphics[height = 0.2\textheight]{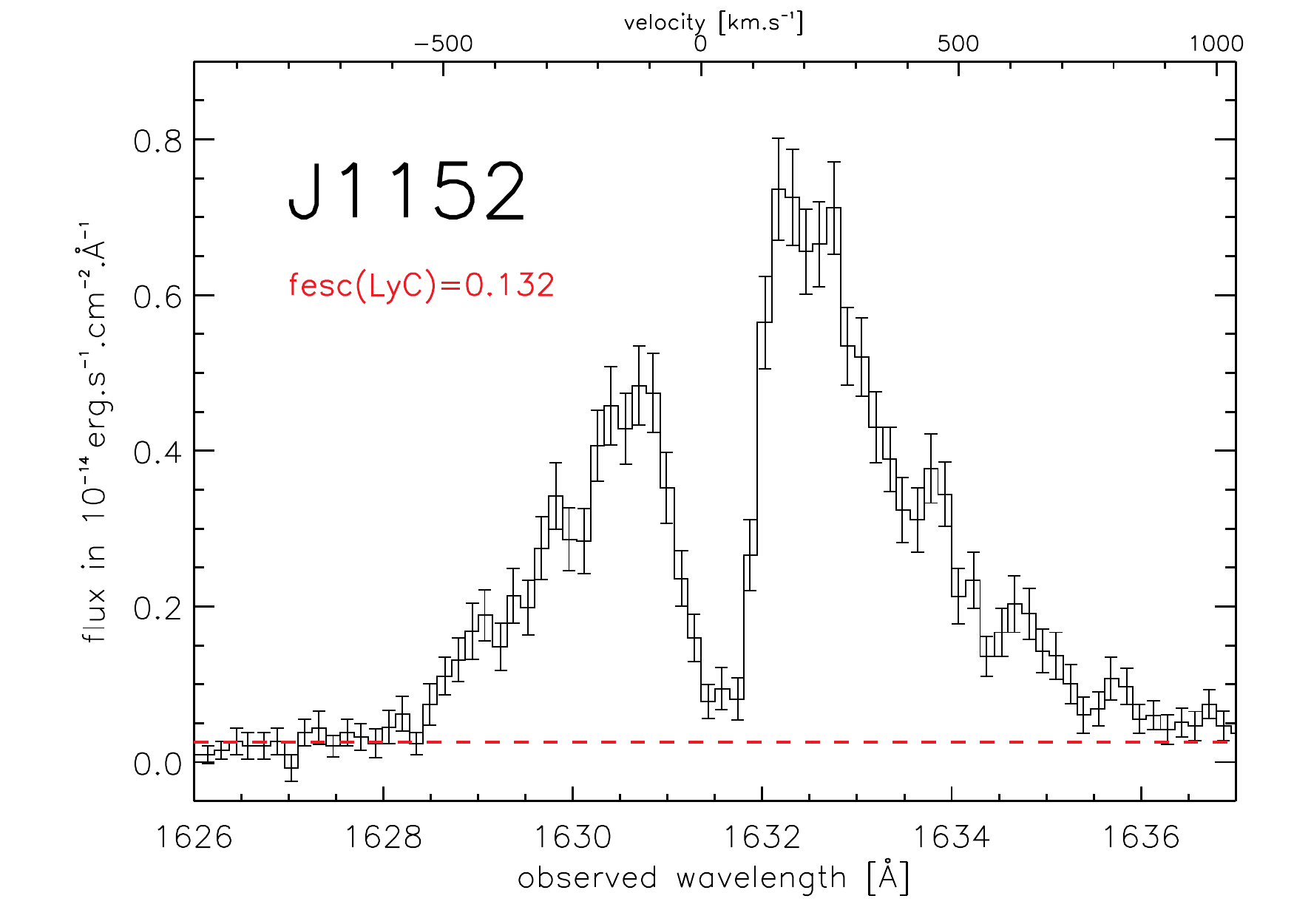} \\
\includegraphics[height = 0.2\textheight]{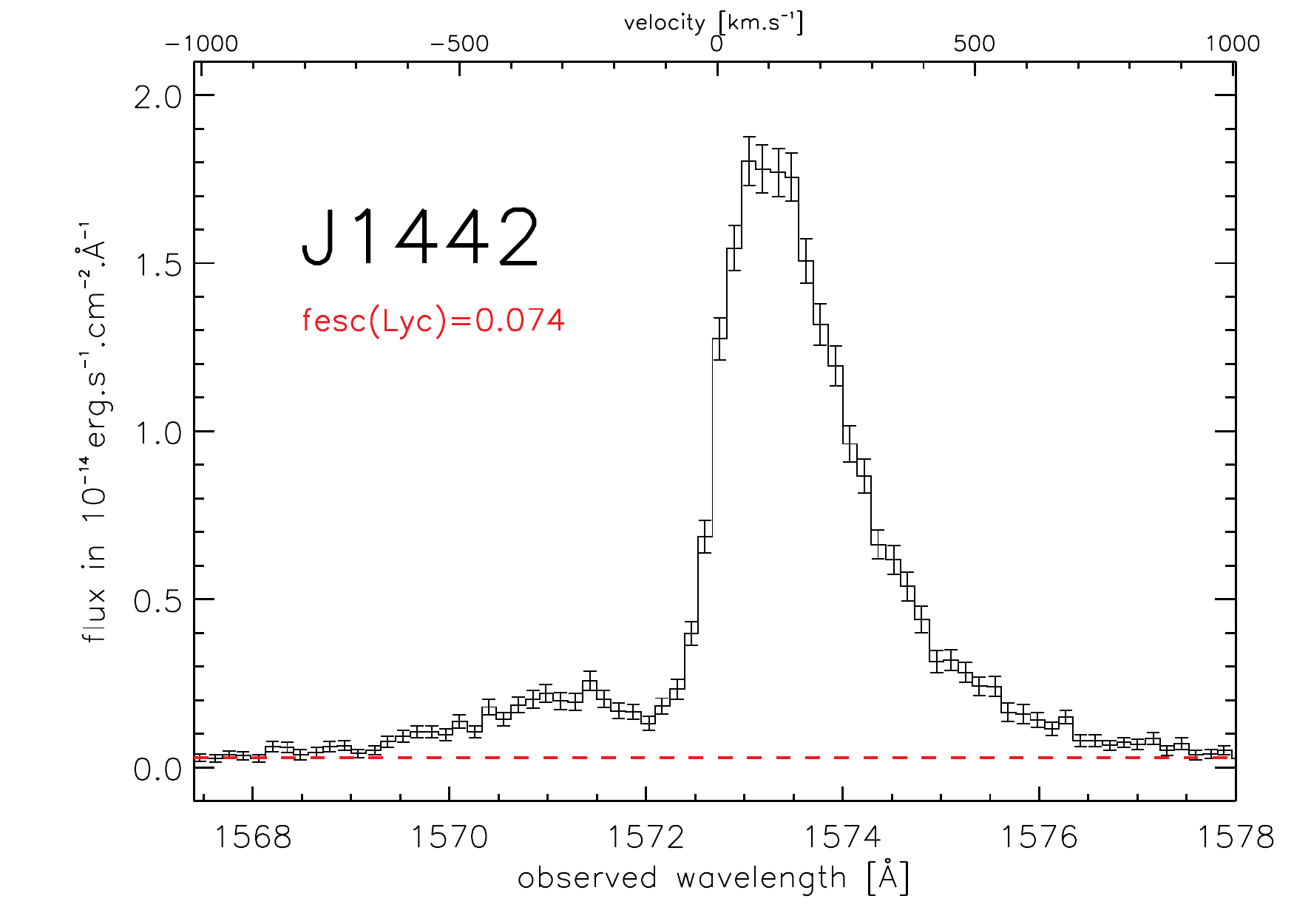} 
\includegraphics[height = 0.2\textheight]{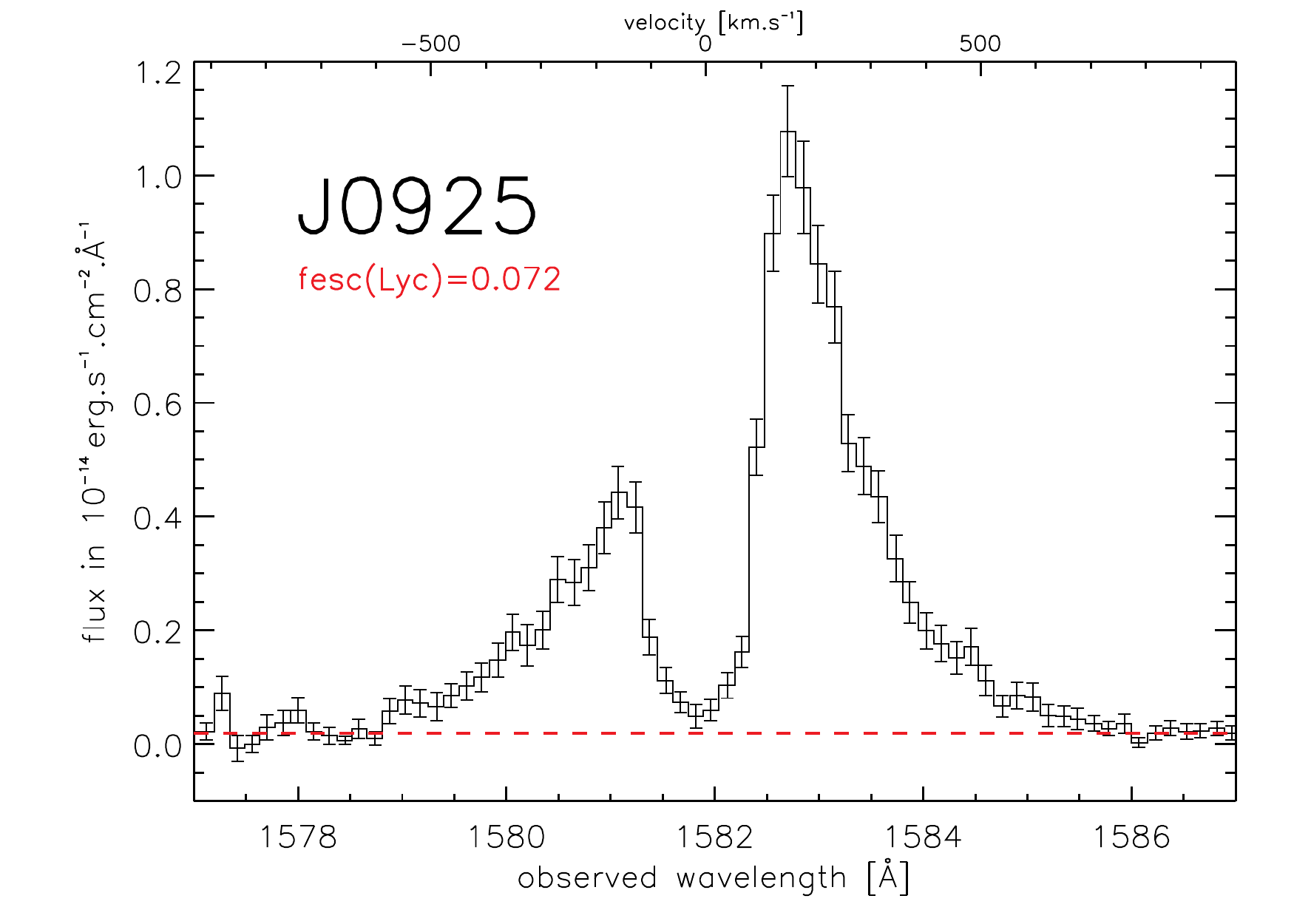} \\
\caption{\lya\ line shapes of LCEs. The whole G160M \lya\ profiles are shown over a small wavelength range, ordered by decreasing \fesclyc\ from top to bottom. They all have very peculiar double peaked profiles, with small peak separation. The dashed lines show the continuum level determined from the entire G160M spectra.
}
\label{spectra}
\end{figure}

\begin{figure}[htb]
\includegraphics[height = 0.2\textheight]{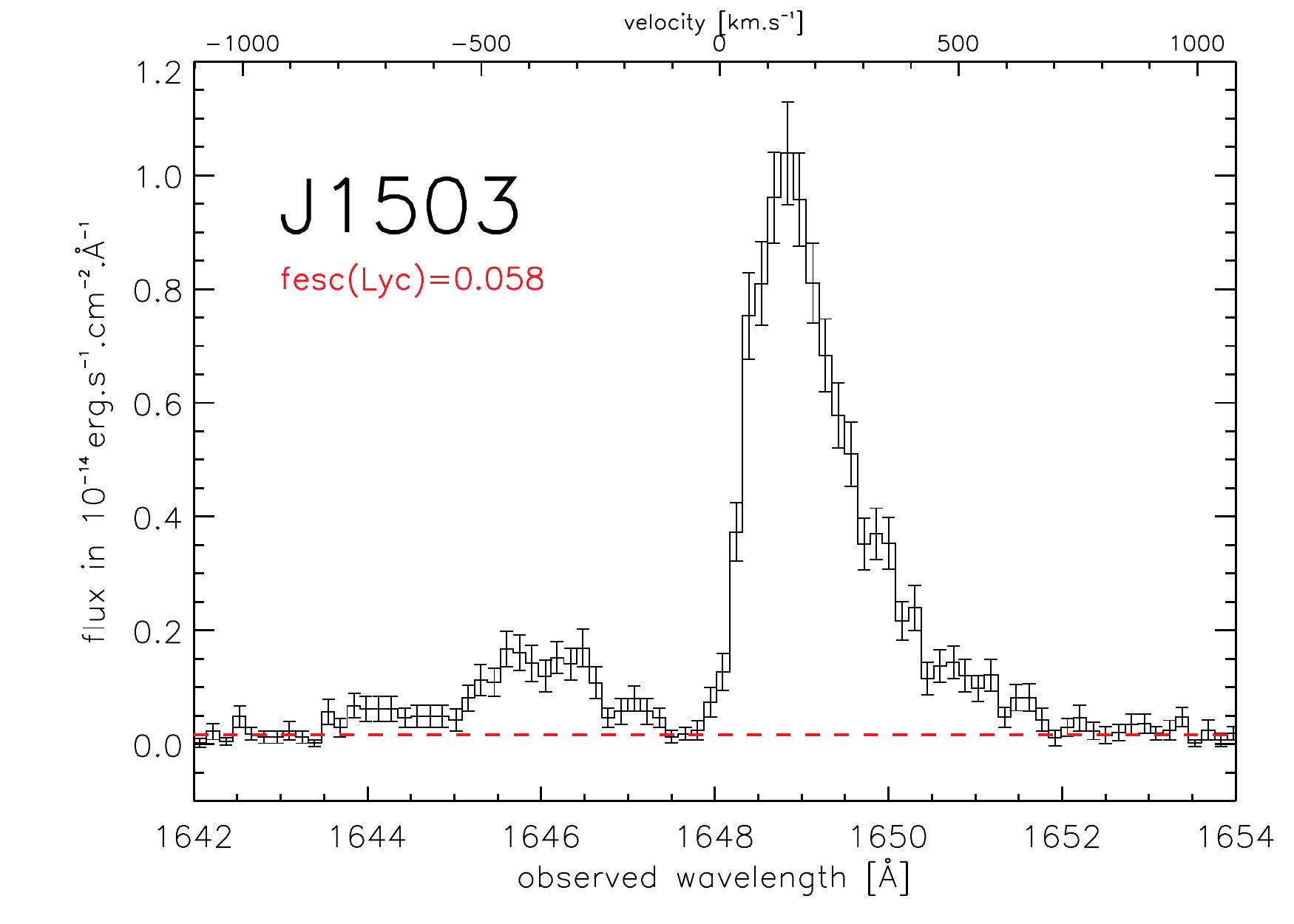} \\
\includegraphics[height = 0.2\textheight]{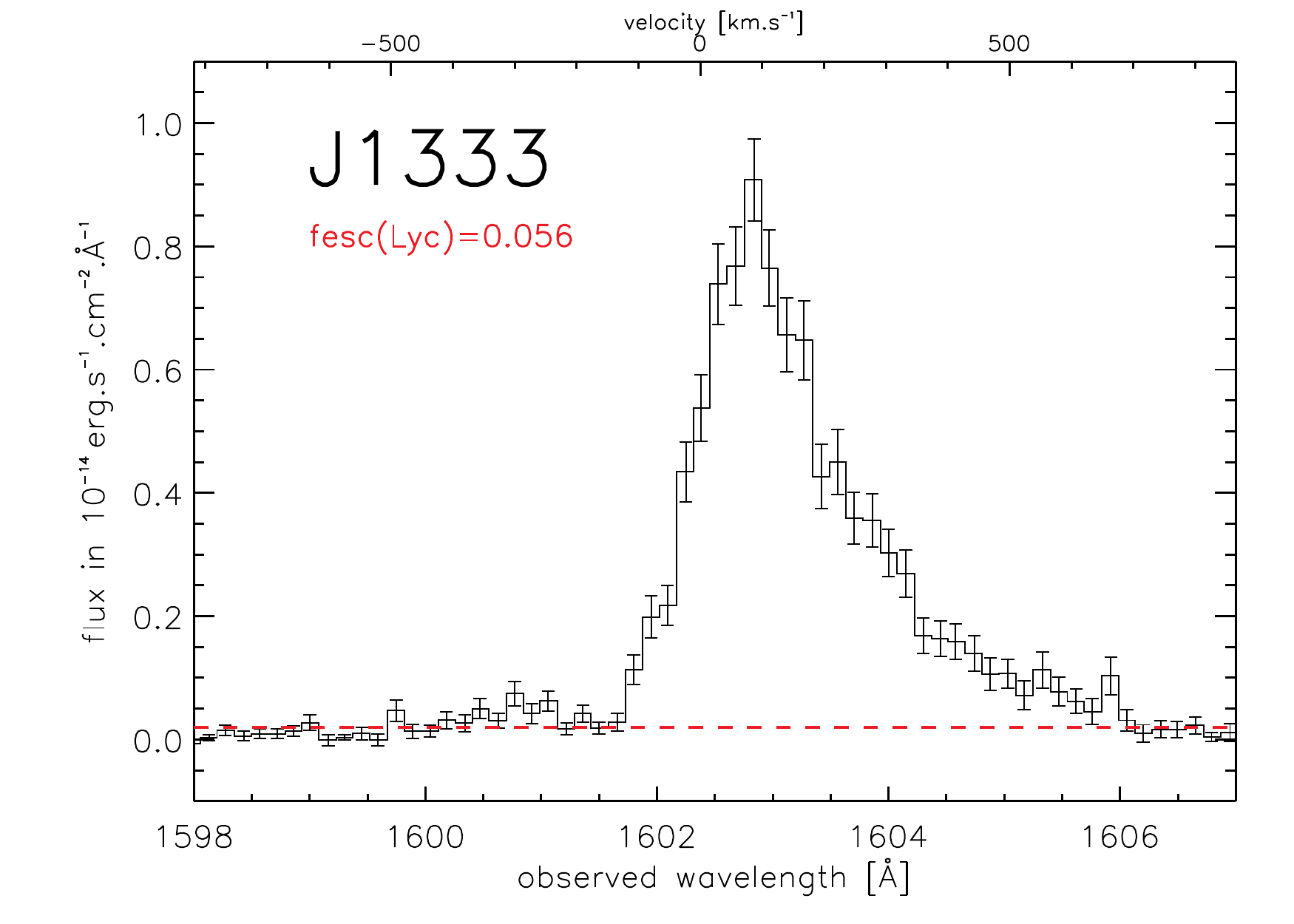} \\
\caption{Same as Fig.\ \ref{spectra} for the remaining two sources.}
\label{spectra2}
\end{figure}

\section{LCEs have narrow double-peaked \lya\ profile}
\label{s_profiles}

In Fig.~\ref{spectra} and  \ref{spectra2} we present the \lya\ line profiles of the five strong LCEs, ordered from top to bottom by decreasing \fesclyc. Once again, the fact that they are all strong \lya\ emitters is very different from the typical low-redshift \lya\ profiles from galaxies studied so far \citep{Wofford13,Rivera15}, except for some of the GPs \citep{Jaskot14, Henry15} and some LBAs \citep{Heckman11, Alexandroff15}. Our objects show two remarkable features, illustrated in the figure. First, it shows strong and narrow lines which are all ``double-peaked", by which we mean \lya\ profiles with two maxima, on each side of the systemic redshift. Second, the profile of the strong leakers never falls below the continuum level. We now quantify and discuss these features and compare them to other objects.

\subsection{Properties of the double-peaked profiles}

As just mentioned, all \lya\ profiles shown here exhibit two peaks, one blueward and one redward of the systemic redshift. This feature, more specifically the ratio between the EW of the blue component to the red one (col.\ 7 in Table~\ref{table_LyaLyC}), has been proposed as an empirical diagnostics for LyC leakage \citep{Erb14, Alexandroff15}. The presence of a blue peak can be understood as a low column density effect. The classical \lya\ P-Cygni profiles emerging from galaxies present a blueshifted absorption, similar to the absorption profiles of Low Ionisation State (LIS) metallic lines. This is due to outflowing neutral gas along the line of sight: the blue photons are seen at resonance for this scattering medium, they are absorbed in the blue and scattered away from resonance (the red peak emerges because most photons are already emitted to the red for the scattering medium). It seems that in LCEs, it is possible to escape when having a blue frequency, meaning that the optical depth of the intervening medium is low enough to allow their escape. We can think of two different scenarios: in a homogeneous medium, both the column density and the velocity of the neutral gas have to be low to shape double peaks with small separations, but a clumpy outflowing medium can also shape double peaks, as proposed and studied in detail in \citet{Gronke16}. To get a clue on the geometry and the kinematics of the intervening neutral gas, we are currently investigating the LIS metallic absorption profiles (Orlitova et al., in prep.).

\begin{figure*}[htb]
\begin{tabular}{cc}
\includegraphics[width=0.45\textwidth]{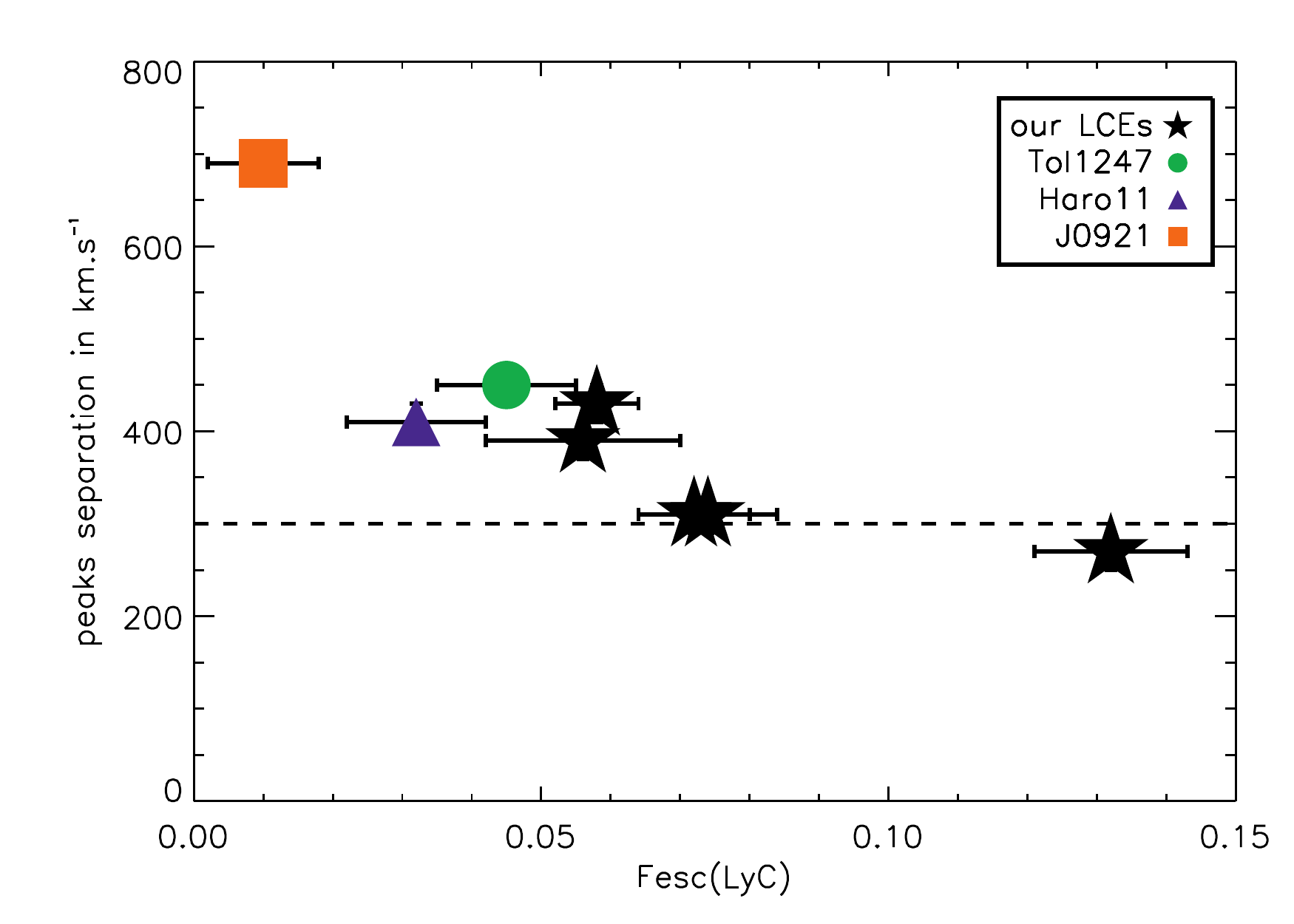} &
\includegraphics[width=0.45\textwidth]{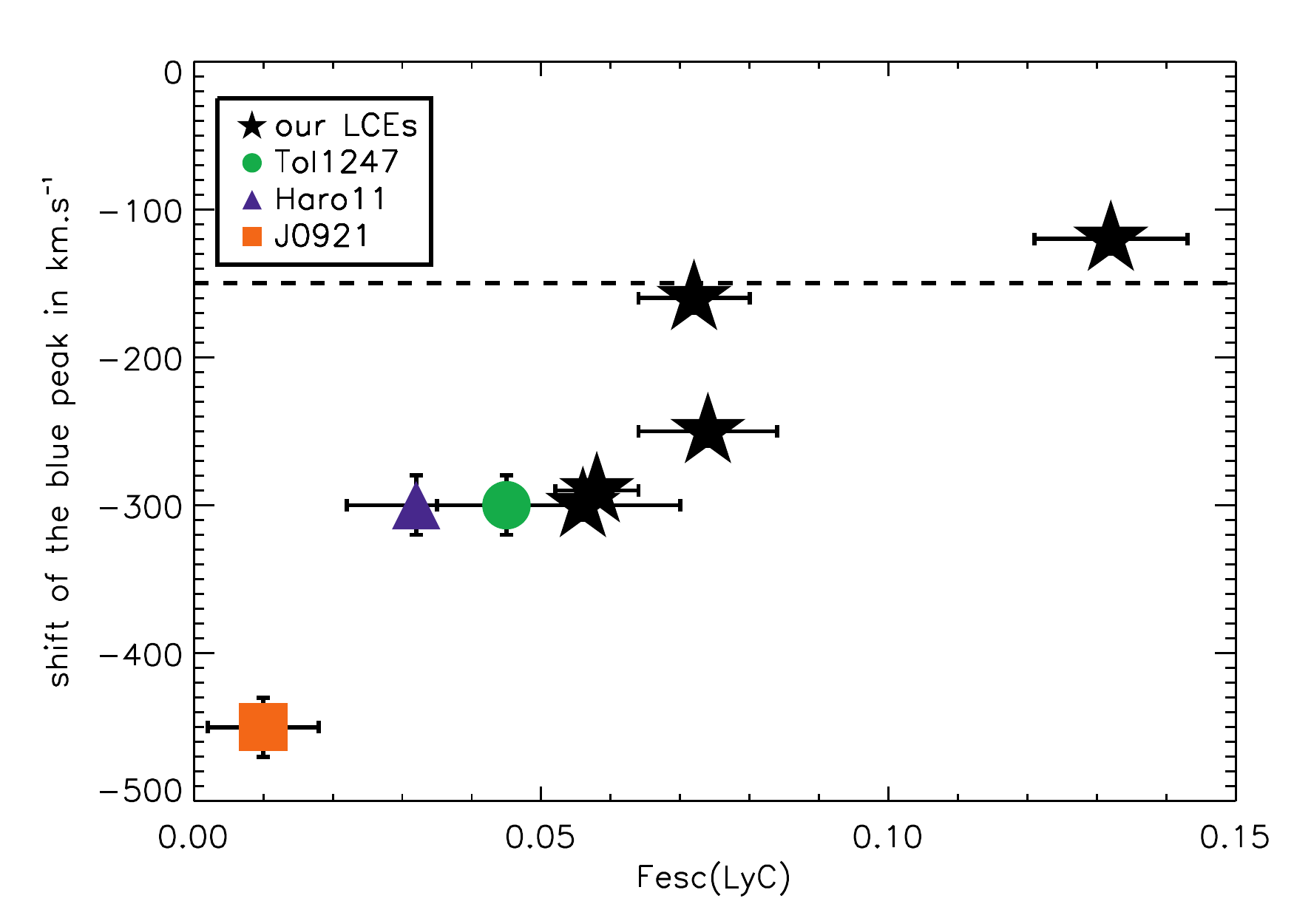} \\ 
\includegraphics[width=0.45\textwidth]{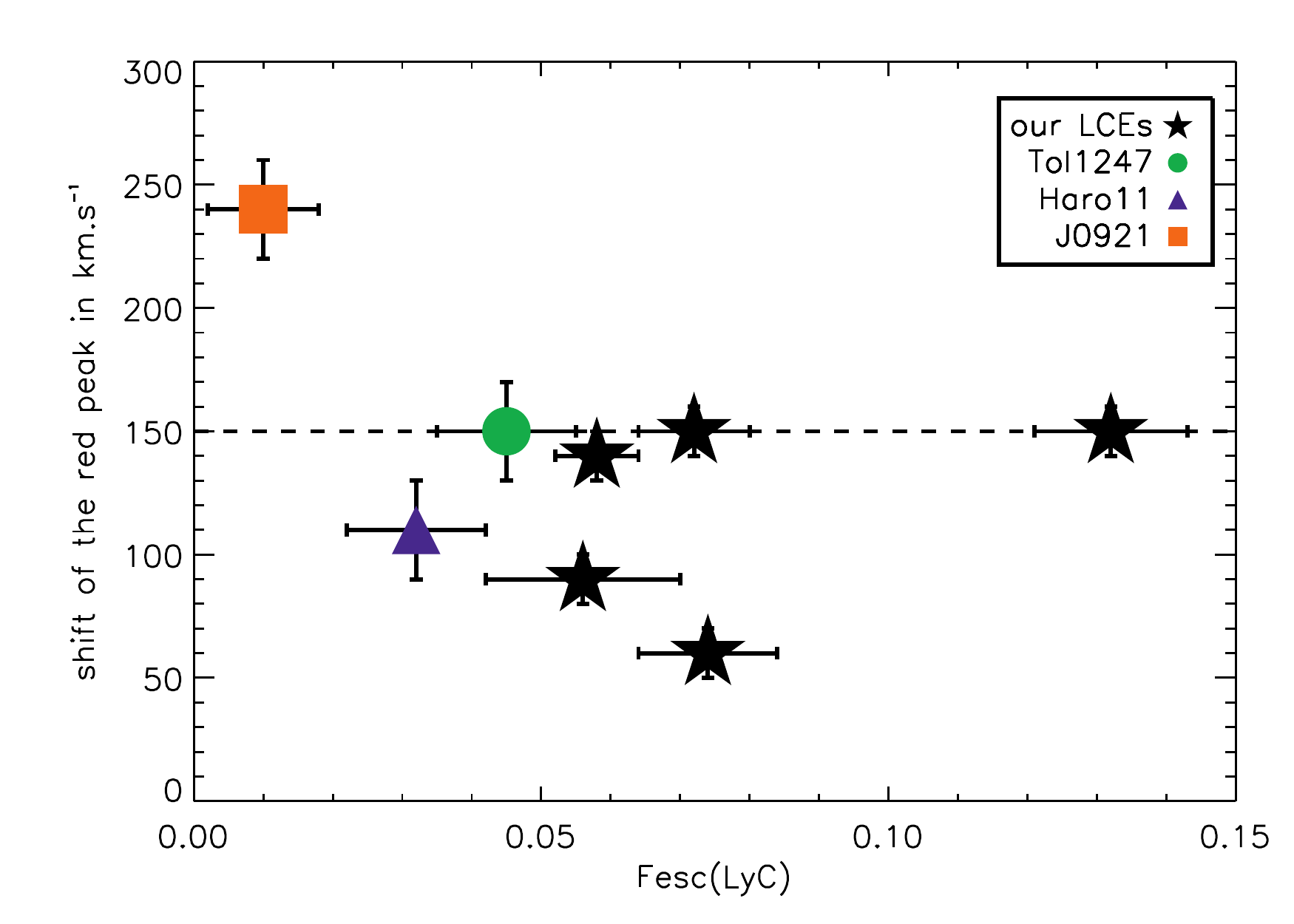} &
\includegraphics[width=0.45\textwidth]{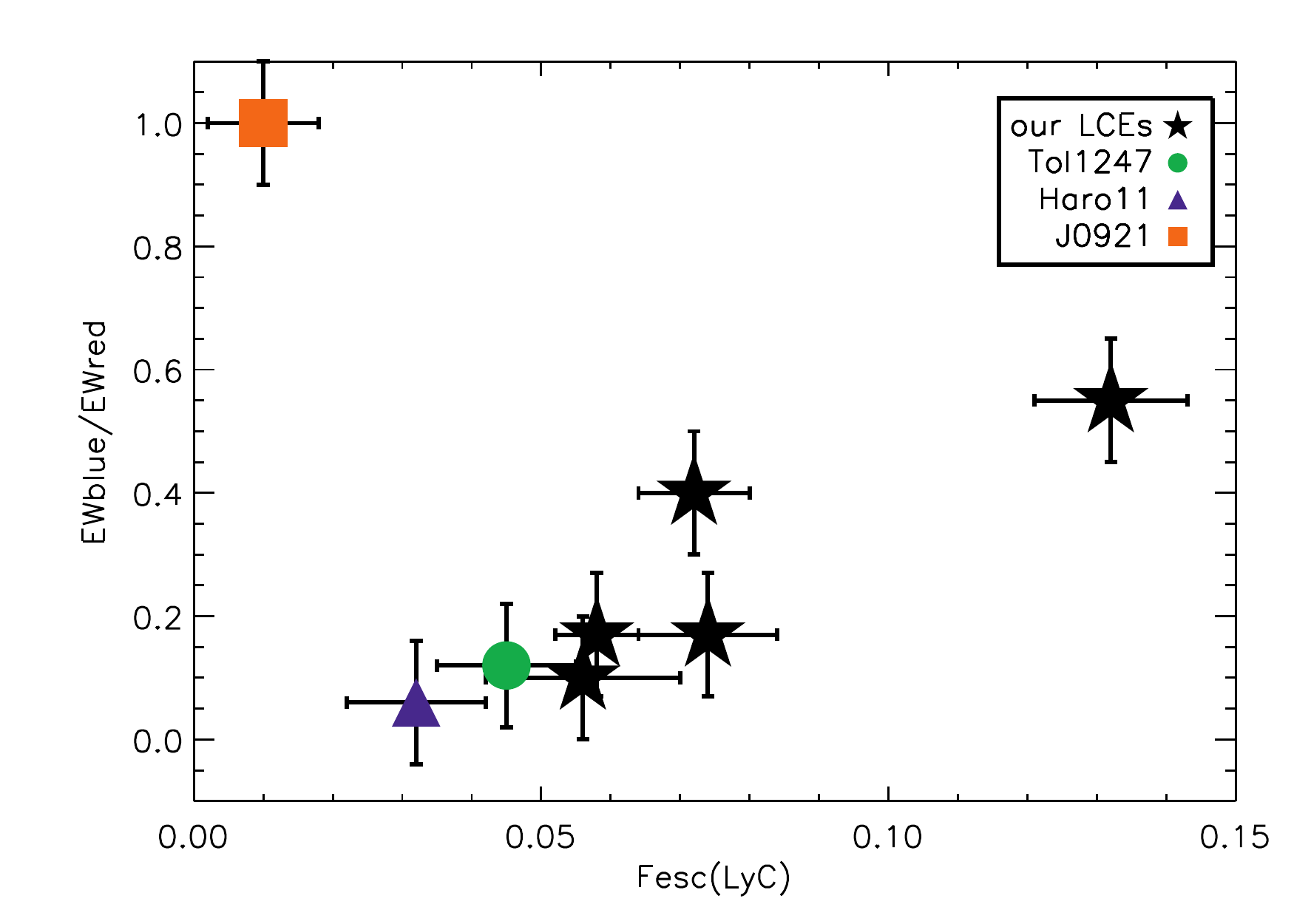} \\
\end{tabular}
\caption{Comparison of the \lya\ diagnostics for LyC leakage with the escape fraction of ionising photons. The correlation between the peak separation and the escape fraction of ionising photons is driven by the shift of the blue peak back to line center. The dashed lines correspond to the theoretical predictions for \fesclyc $\sim 30\%$ ($\tau_{\mathrm{LyC}} \sim 1$) from \cite{Verhamme15}}
\label{peaks_vs_fescLyC}
\end{figure*}

In Fig.~\ref{peaks_vs_fescLyC} we present the peak shift measurements versus the escape fraction of ionising photons for all objects of Table \ref{table_LyaLyC}.
A clear anti-correlation is found between the escape fraction of ionising photons and the separation of the peaks in the \lya\ profile (first panel) decreasing to $\vsep \sim 300$ \kms\ for the highest \fesclyc. 
Along the same lines, \citet{Hashimoto15} showed that the peaks separation is lower, and the velocity shift of the red peak is smaller, on a sample of $z \sim 2.2$ LAEs than typically measured in LBGs \citep{Shapley03,Kulas12}, in correlation with lower \nh\ estimates for LAEs than for LBGs, from \lya\ line profile fitting \citep[][their Fig 11]{Hashimoto15}. 
As predicted by radiation transfer modelling in idealised geometries \citep{Verhamme15}, the width and the shift of the main peak or the peak separation trace the column density of the scattering medium. In particular, a peak separations of $\sim300$ \kms\ and smaller correspond to \lya\ radiation transfer in media with column densities below $\nh=1.6\times10^{17}$ cm$^{-2}$. Our predictions were made for media with an optical depth of $\tau_{\rm LyC} \sim 1$ for the ionising radiation, corresponding to an escape fraction of $\sim 30\%$, higher than observed in our sample, but these characteristic signatures seem to hold even at lower escape fractions. Indeed, our five LCEs all have small peak separations, although slightly higher than the predicted $300$ \kms\ value, as reported in Table~\ref{table_LyaLyC}.  
Knowing \fesclyc\ for our objects, we can estimate the column density of neutral gas along the line of sight corresponding to their LyC escape fractions:
\begin{equation}
\nh = - \ln($\fesclyc$) / \sigma_0, 
\end{equation}
where $\sigma_0 = 6.3 \times 10^{-18}$ cm$^2$. This corresponds to \nh = $(3.2-4.6) \times 10^{17}$cm$^{-2}$ for our LCEs. So the anti-correlation between \fesclyc\ and \vsep\ disucssed above translate into an anti-correlation between \nh\ and \vsep.

Interestingly in our data, the behaviour of \vsep\ is not driven by the location of the red peak (third panel), but by the blue one (second panel). \citet{Henry15} already reported the strongest correlation between the shift of the blue peak, $V_{\mathrm{peak}}^{\mathrm{blue}}$, and \fesclya\ for GPs, we also find the strongest correlation between $V_{\mathrm{peak}}^{\mathrm{blue}}$ and \fesclyc. In the case of {\it Ion2}, with current lower spectral resolution data, the blue peak is consistent with systemic redshift \citep[see Table 1 in ][]{deBarros16} and \fesclyc\ is very high ($> 50$\%), so {\it Ion2} would also follow the reported trend.
In fact the location of the red peak is very close to or below the theoretical prediction for LyC leakage ($V_{\mathrm{peak}}^{\mathrm{red}} < 150$\kms) and almost the same for 7 out of 8 objects, only the LBA J0921+4509 has clearly higher peak offsets, so no trend is seen between the escape fraction of ionising photons and the red peak location. A (very) small velocity shift of the main peak rather appears as a necessary condition for LyC escape. But the location of the blue peak seems to correlate more strongly with the escape of ionising photons.  
A stronger trend with the blue peak compared to the red peak was not predicted by our models. Further theoretical work has to be done to understand this observational trend, clumpiness appearing as an interesting geometry to explore \citep{Gronke16}.

As already mentioned, the ratio of the EWs of the blue and the red peaks has been proposed as a empirical criterion for LyC escape, because it seems to correlate with the depth of the LIS lines \citep{Erb14, Alexandroff15}.  We investigate this correlation on the fourth panel of Fig~\ref{peaks_vs_fescLyC}, but the trend appears less clear than for the peak separation, or the blue peak shift: our strongest leaker J1152+3400 has a prominent blue peak, but the object with the lower escape fraction, LBA J0921+4509, too. The second strongest leaker, J1442$-$0209, has only a small blue bump, but the whole profile is clearly in emission. There is no clear theoretical support to this criterion: kinematics rather than column density governs the relative heights of the peaks, at least in the homogeneous media studied so far. The peak separation, and no underlying absorption appear to trace LyC escape from galaxies better than the EWb/EWr ratio, although these correlations have to be studied on a bigger sample.

\subsection{Comparison with GPs and LBAs}

\begin{figure*}[htb]
\begin{tabular}{cc}
\includegraphics[width=0.45\textwidth]{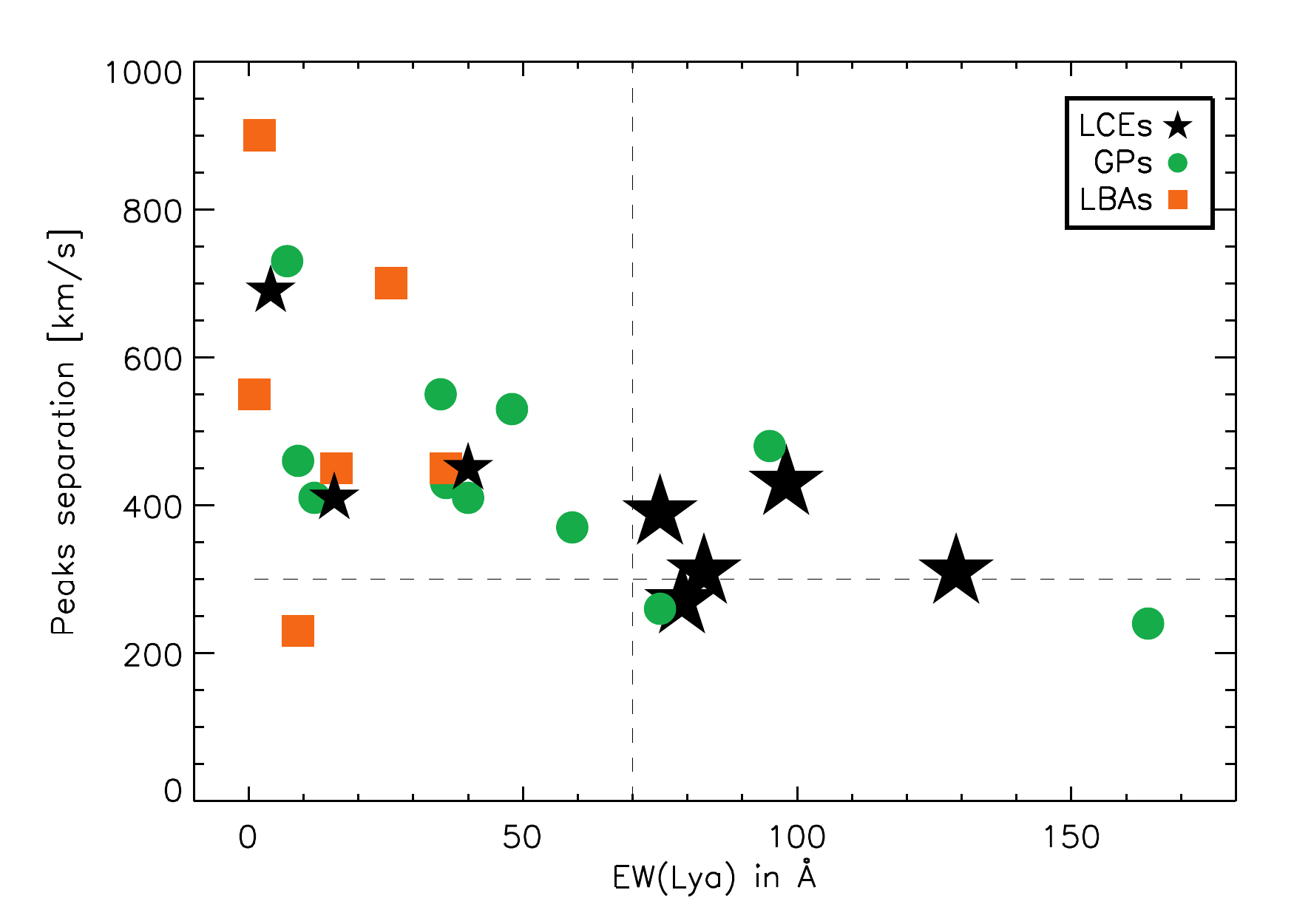} &
\includegraphics[width=0.45\textwidth]{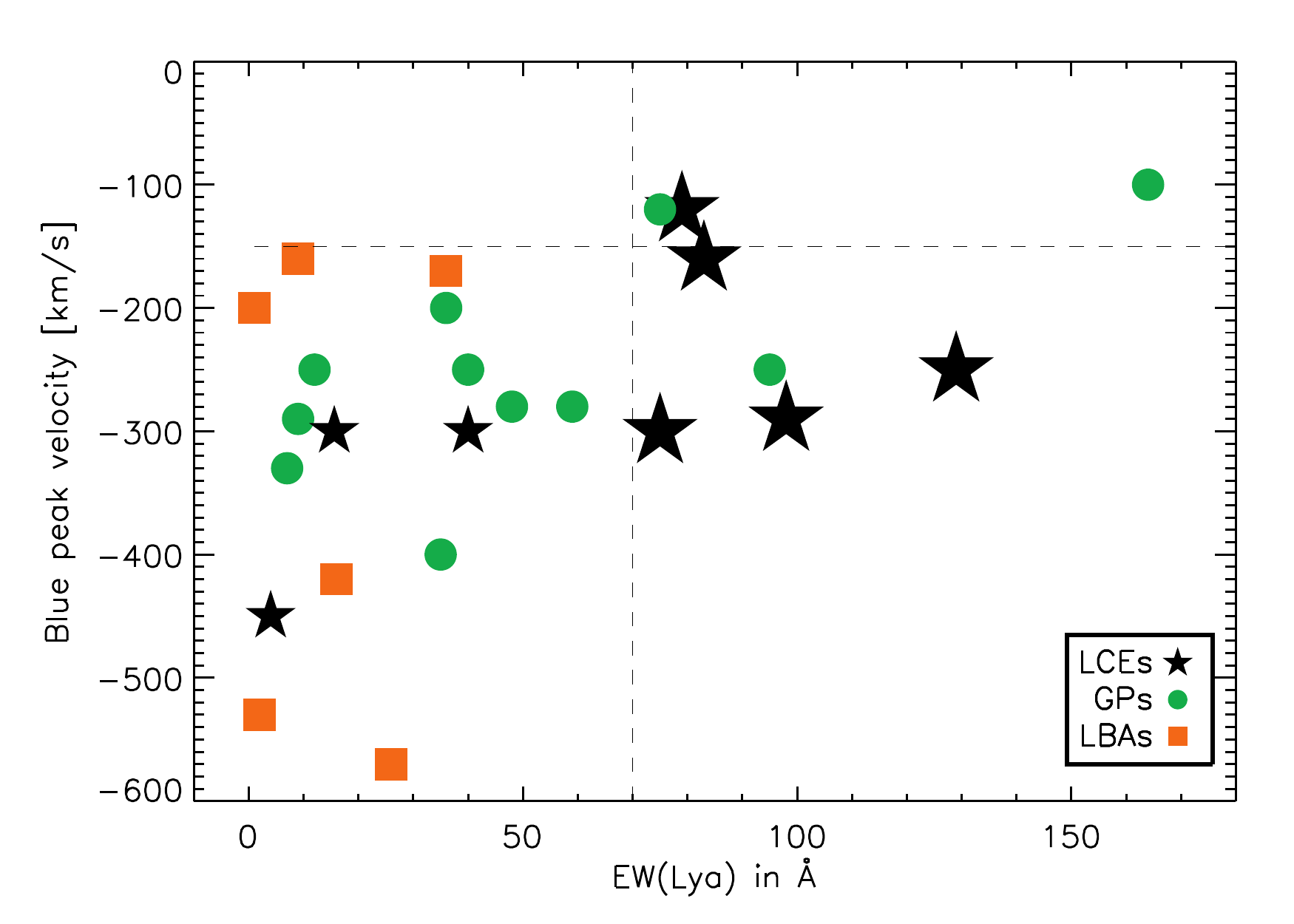}\\ 
\includegraphics[width=0.45\textwidth]{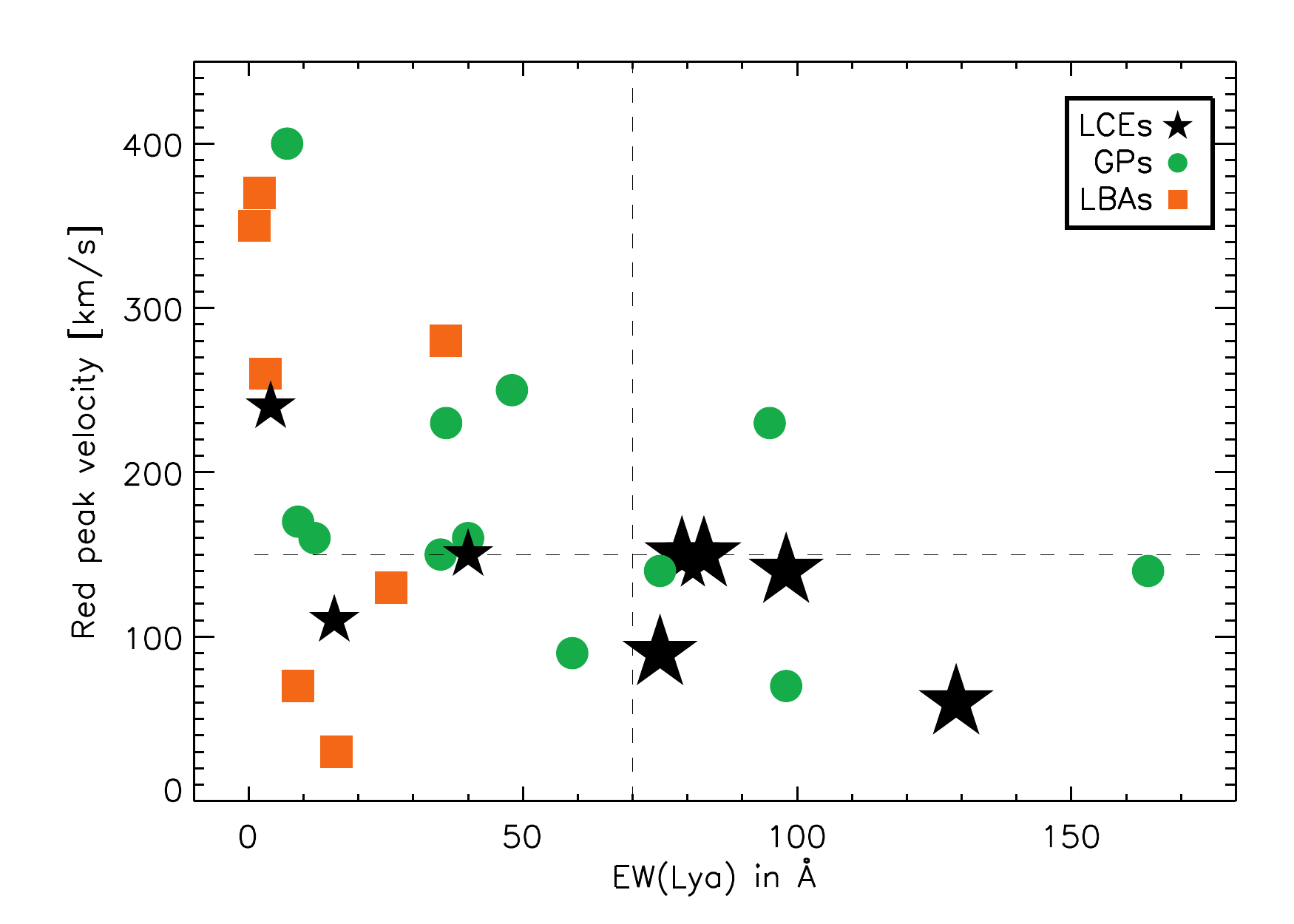} &
\includegraphics[width=0.45\textwidth]{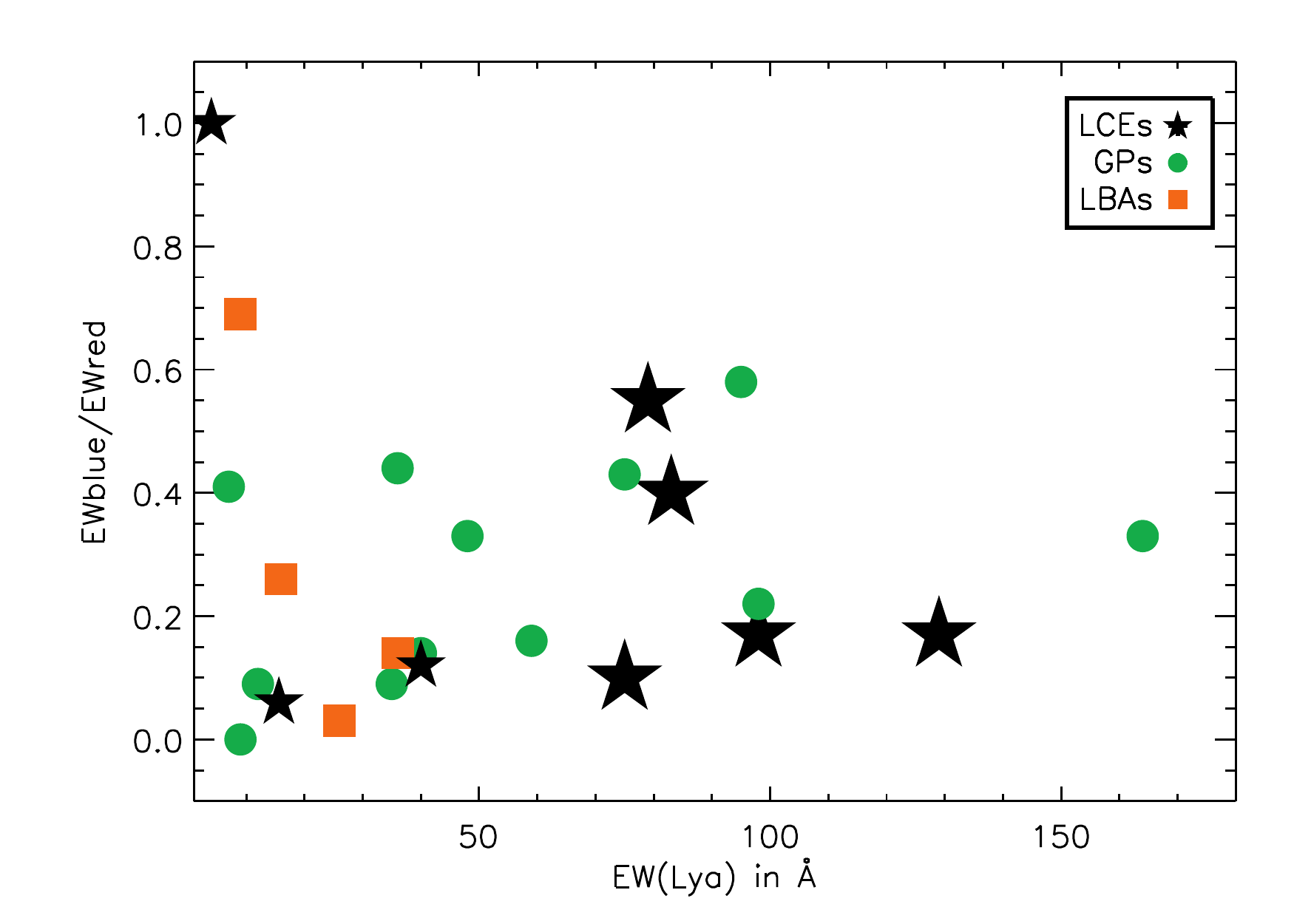} \\
\end{tabular}
\caption{Comparison of the \lya\ diagnostics for LyC leakage over the comparison samples of LBAs and GPs with the same measurements on LCEs. The dashed lines correspond to the theoretical predictions for \fesclyc $\sim 30\%$ ($\tau_{\mathrm{LyC}} \sim 1$) from \cite{Verhamme15}.
Black stars, green circles, and orange squares show the data of the LCE, GP, and LBA samples, respectively.
}
\label{peaks_vs_EWLya}
\end{figure*}

Finally, we examine the various quantities characterising the \lya\ profiles as a function of the \lya\ equivalent width, which allows us to include also the comparison samples. The results are presented in Fig~\ref{peaks_vs_EWLya}. For all samples we see the same trends between the peak separations and EW(\lya), and between the blue peak location and  EW(\lya), in agreement with the correlations presented above: EW(\lya) vs \fesclyc\ (Fig.~\ref{LyaStrength_vs_LyC} left), and peak separation/blue peak location vs \fesclyc\ (Fig.~\ref{peaks_vs_fescLyC}). 
Only objects with high EWs have small peak separations \vsep.
Although no direct measurements of \fesclyc\ are available for the comparison sample, some of the objects share all \lya\ characteristics of our LCEs: a strong EW(\lya) ($>70$\AA), a high \fesclya ($>20$\%), a small peak separation, and the whole \lya\ profile in emission. This comparison suggests that a fraction of the GP and LBA samples shown here may also be strong (\fesclyc$>5\%$) LCEs.

\section{Discussion}
\label{s_discuss}

\subsection{The lack of underlying absorption}

\begin{figure}
\includegraphics[width = 0.5\textwidth]{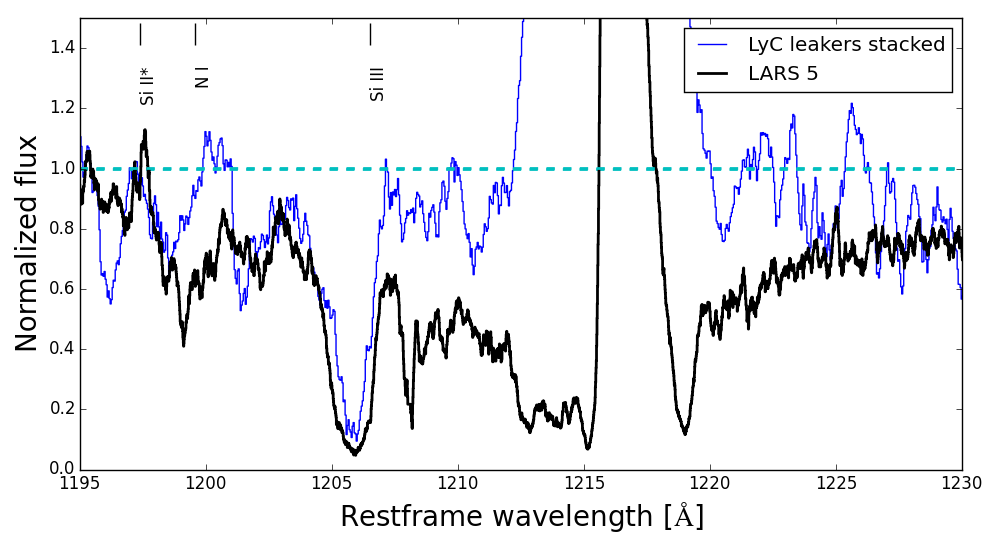} \\
\caption{Comparison of the amount of underlying absorption below the \lya\ emission between a stacked spectrum of our five LCEs (thin blue curve) and LARS05 (thick black curve), chosen as representative of typical \lya\ profiles in local galaxies: the leakers have clearly less underlying absorption. See text for more details.}
\label{continuum}
\end{figure}

As shown in Fig.~\ref{continuum}, a stack of the \lya\ profiles of our LCEs over a broad wavelength range (thin blue curve) shows no/weak underlying absorption.  This is a very unusual feature. LARS galaxies, selected for their strong H$\alpha$ to trace recent star-formation, have often \lya\ in absorption, and when in emission, it is usually broad, redshifted, asymmetric, and lying on top of a broad absorption feature \citep[see also][]{Wofford13}. 
As an illustration, we overplot the \lya\ spectrum of LARS05 (thick black curve), a strong LAE studied in details in \citet{Duval16}. The underlying absorption is clearly visible for this object, especially in the red wing. LARS05 is a galaxy where we do not expect LyC leakage: in \citet{Duval16}, they fit the underlying absorption trough and estimate $\nh \sim 4\times 10^{20}$ cm$^{-2}$.  We also notice that some objects presented in \citet[][Fig 8]{Alexandroff15}, e.g. J1416+1223 or J1414+0540, present a broad absorption feature, they are probably not good candidates for LyC escape. The LCE from \citet{Borthakur14} discussed above, J0921+4509, is presented in the same figure, and its underlying absorption is less pronounced. Our LCEs resemble more closely the objects with no underlying absorption as e.g. J0926+4427 (which is also the GP LARS14).

The origin of the broad absorption around \lya\ {\bf can be either stellar or interstellar}. 
For starbursts with ages above $\sim10$ Myr, or constant star-formation history, synthetic stellar spectra show an absorption around \lya\ \citep[e.g.\ Fig.\ 2 in][]{Schaerer08, Pena13}. The fact that we don't see this feature in any of our LCEs is {\bf consistent with} the very young age of their bursts, as determined by SED fitting \citep{Izotov16b}. But it also means that the interstellar medium is transparent enough to not imprint any visible wing of the Lorentzian profile. 

The underlying absorption seen in most \lya\ spectra \citep[e.g.\ LARS and eLARS, or Figs.\ 4-9 in][]{Wofford13} is often broader that the stellar absorption, it is likely due to interstellar absorption: the spectrum emerging from a star-forming region, i.e. the UV stellar continuum plus a nebular \lya\ recombination line, is absorbed by the interstellar gas along the line of sight, imprinting a Voigt profile on the transmitted\footnote{photons escaping the medium without scattering constitute the transmitted spectrum} stellar UV continuum. This profile usually has very broad wings for typical column densities of the ISM ($\sim\pm 6000$ \kms\ for $\nh \sim 10^{21}$ cm$^{-2}$), it is centered on the bulk velocity of the scattering medium. The central part of the trough on the transmitted spectrum is partially refilled by the resonant \lya\ emission, but the spectral shift/broadening due to radiation transfer through the ISM is usually smaller than the width of the Voigt profile, and the wings are still visible on the emergent spectrum. Our spectra do not show any sign of absorption below the \lya\ line, except for J1333+6246 which may have some absorption bluewards. The absence of the absorption is a model independent indication for low column density of neutral gas in front of the UV source, along the line of sight: the Voigt profiles of the absorbing gas along the line of sight are narrower than the wings of the scattered emission components. As discussed above, \fesclyc\ correspond to \nh = $(3.2-4.6) \times 10^{17}$cm$^{-2}$ for our LCEs.

The presence of an underlying absorption in the \lya\ spectra of typical star-forming galaxies of the local Universe is certainly enhanced by an aperture effect. LBAs, LARS and especially eLARS galaxies are closer in redshift, and spatially more extended than typical GPs and our compact LCEs. The COS aperture captures only partially the \lya\ emission from these galaxies, as well demonstrated in \cite{Hayes13, Hayes14}. The transmitted \lya\ spectrum is the same as long as the source is inside the COS aperture, but the scattered component will escape the galaxy over a broad area, which extends over the COS aperture for local starbursts, but may still fall into the detector for these very compact sources.  

In summary, our LCEs do not show the usual broad absorption below the \lya\ line, probably for a combination of the three reasons proposed above: their bursts are very young, still in the regime where O stars dominate the stellar population, so no stellar absorption; their ISM is very transparent along the line of sight since ionising radiation escapes, so no additional interstellar absorption; they are so compact that probably most of the scattered component of the \lya\ emission is recovered inside the COS aperture, refilling absorption, if any. More data are needed to determine if these properties are universal of LCEs: are all bursts in LCEs especially young and compact? Can we find some objects emitting ionising radiation, and with a detectable broad absorption component inside the \lya\ emission? Or can we use the presence/absence of underlying absorption as an independent probe of LyC leakage?  One LyC detection has been reported so far from a $z\sim.83$ galaxy with \lya\ in absorption \citep{Vanzella10}, which is unexpected from radiation transfer modelling, and given the results of the actual study. If the LyC detection is confirmed further in this interesting object, it will become a great challenge for further theoretical investigations.

\subsection{Applicability to {\bf high-redshift} observations}

\subsubsection{Effect of a lower spectral resolution}

\begin{figure}
 \includegraphics[width = 0.5\textwidth]{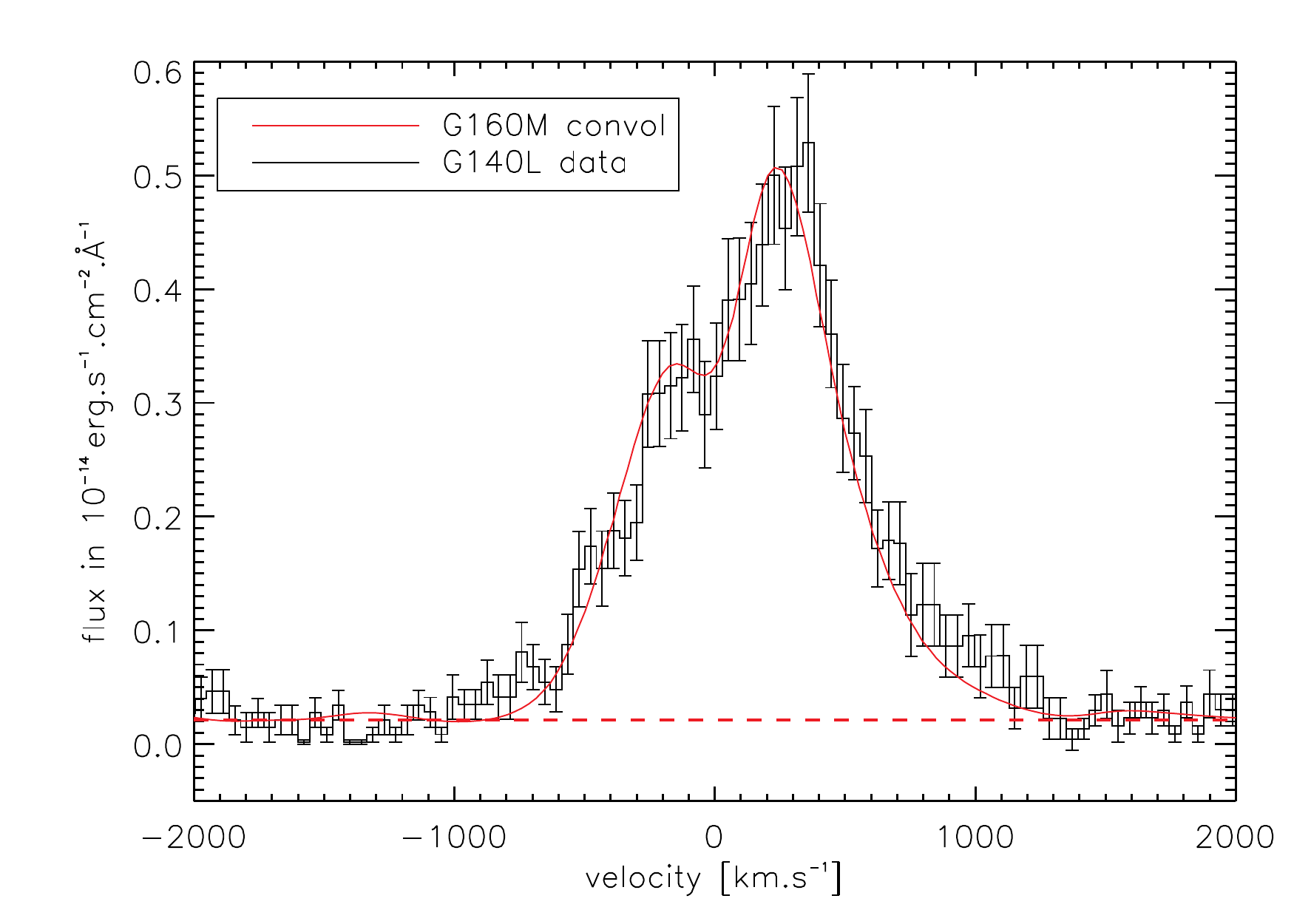} 
\caption{\lya\ spectrum of J1152+3400 in the G140L spectrum, $R \sim 1000$, comparable to high-redshift surveys data for LAEs. 
}
\label{low_res}
\end{figure}

\begin{figure*}[htb]
\begin{tabular}{cc}
\includegraphics[width=0.45\textwidth]{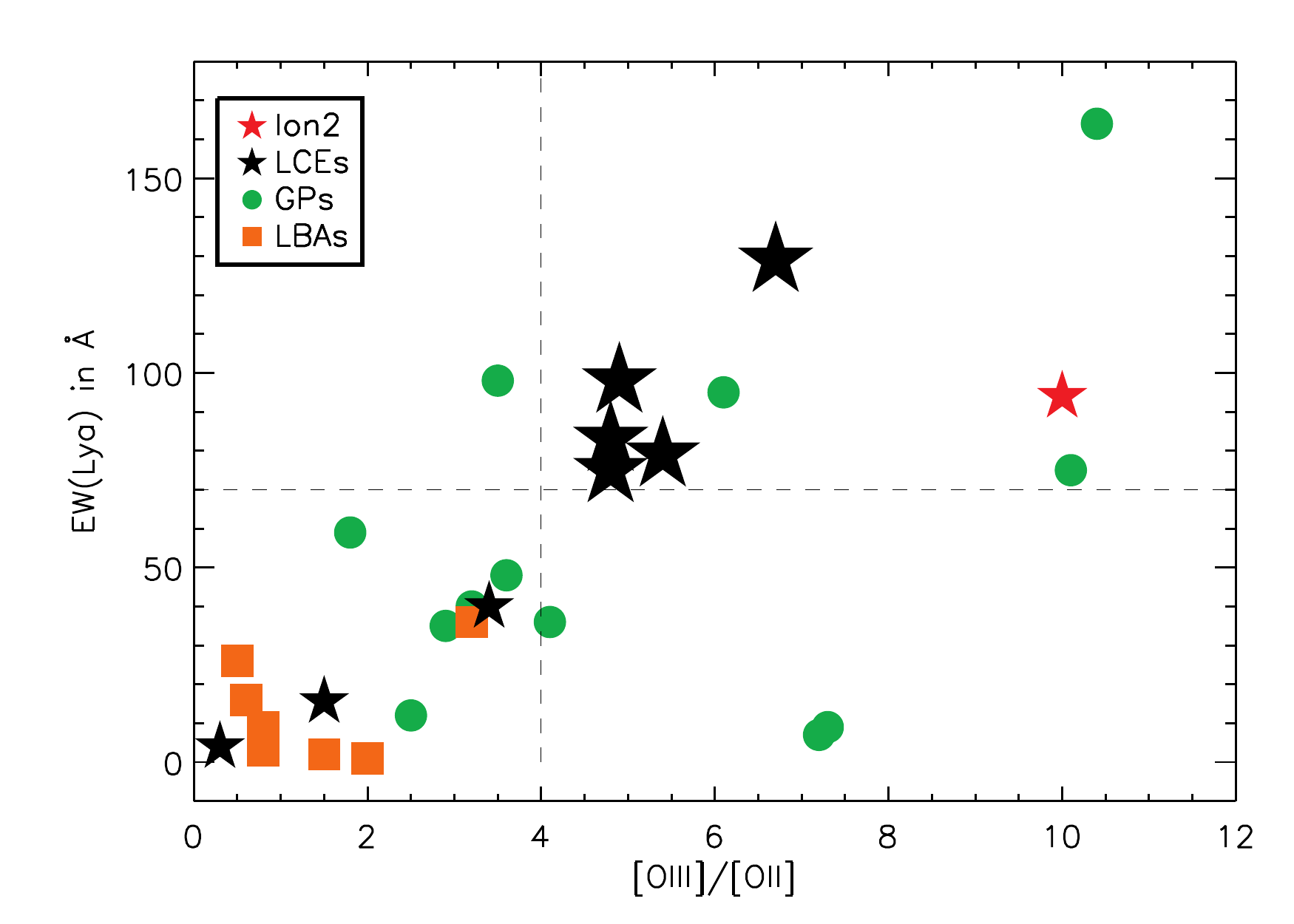} &
\includegraphics[width=0.45\textwidth]{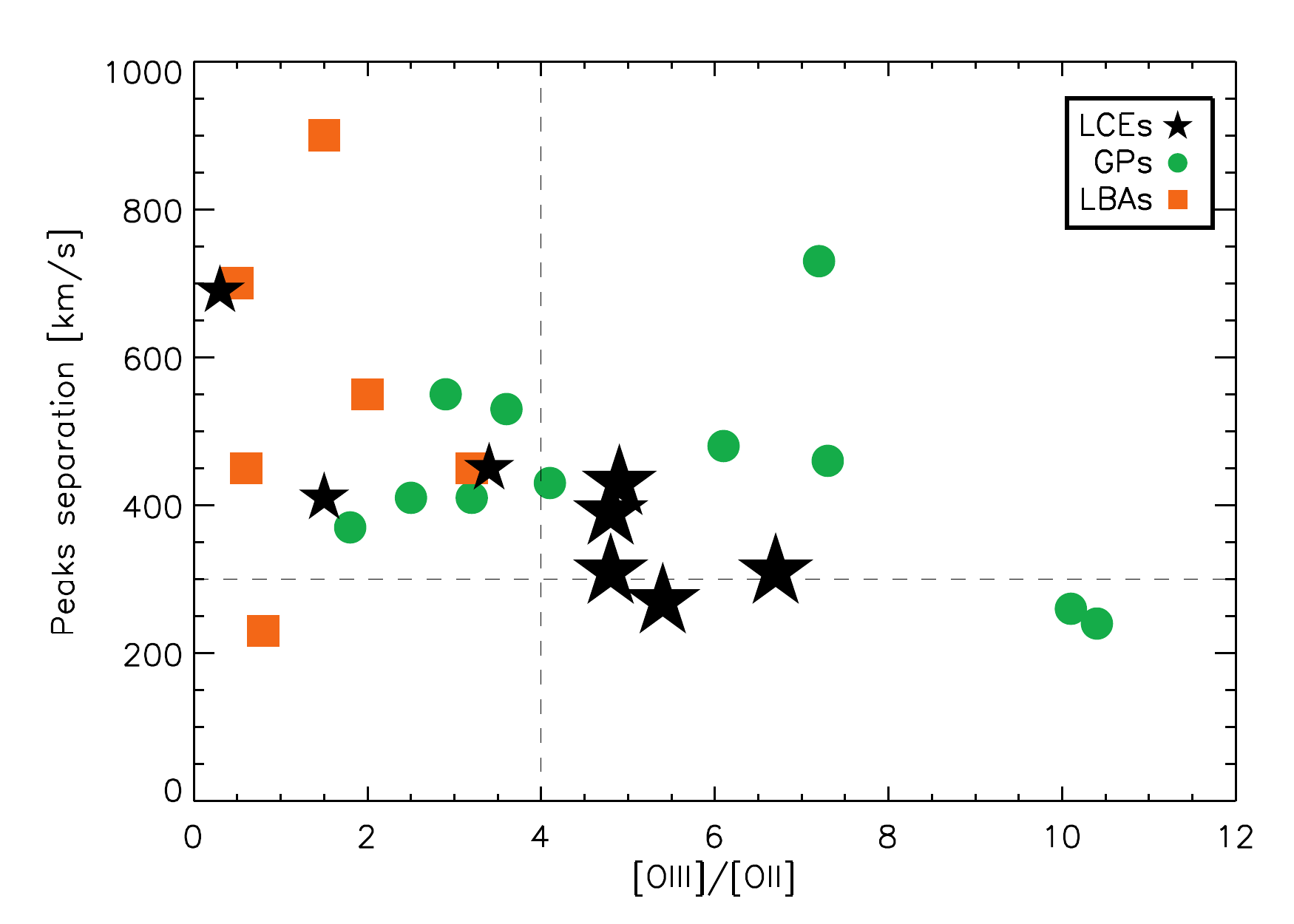} \\
\end{tabular}
\caption{Comparison of the \lya\ diagnostics for LyC leakage with \oiii/\oii\ ratios.}
\label{lya_vs_O32}
\end{figure*}

With the exception of few lensed galaxies at $z \sim 3$ observed with $R \sim 5000$ \citep{Fosbury03,Vanzella16b}, \lya\ spectra of high-redshift galaxies do not reach the medium resolution of the {\sl HST}/COS G160M grisms. For illustration, we show in Fig.~\ref{low_res} the G140L \lya\ profile of our strongest leaker, J1152+3400, in black. We checked that when we degrade the G160M spectrum -- with an effective resolution of $R \sim 4000$ -- 
to match qualitatively the effective spectral resolution of G140L for this object, which is estimated at $R \sim 1000$, we find a similar alteration of the profile as shown by the red curve in Fig.~\ref{low_res}.
The dip between the peaks is not visible anymore on the low resolution data, the peaks are barely resolved. Indeed, the peak separation is $\vsep = 270 $\kms\ for this object, just below $\Delta V \sim 300 $\kms\ corresponding to a resolving power of $R=1000$. 
We tested on the G140L data how the EWs and velocity shifts are altered by low spectral resolution. We did the same measurements on the G140L spectra of our five leakers, and found that we recover EW(\lya) of the G160M measurements within the uncertainties of the continuum level estimation, but the velocity shifts are more strongly affected, up to hundreds of \kms\ depending on the spectral shape. 
This implies that a spectral resolution of $R \ga 1000$ is needed to measure the smallest peak separation, typical of strong LCEs: the \lya\ spectra of LCEs in these surveys should appear as strong and narrow single peaked spectra. If the double peaks are resolved, it means that \vsep\ is too large to trace strong LyC escape. 

\subsubsection{Effect of IGM attenuation}

The neutral fraction of the IGM increases with redshift \citep{Madau95} and alters the transmission of high-redshift \lya\ spectra \citep[e.g.][]{Haiman02, Santos04, Dijkstra14}. For example, \citet{Laursen11} showed how the blue part of the \lya\ line can be scattered out of the line of sight of high redshift galaxies, by neutral gas along the way, in hydrodynamical simulations of galaxy formation, altering the synthetic \lya\ profile shapes. Observations also witness a sudden transition in the evolution of the \lya\ properties of galaxies with redshift: the fraction of \lya\ emitting galaxies among LBGs increases regularly with redshift, up to $z \sim 6$ where a sharp decrease is observed \citep[e.g.][]{Schenker14}. The same behaviour has been reported for the evolution of \fesclya\ with redshift \citep{Hayes11}. However IGM attenuation is a very stochastic process, and greatly varies from line of sight to line of sight \citep{Inoue14}. Maybe for that reason, a significant fraction of double-peaked \lya\ emitters has still been observed at intermediate redshifts ($z \sim 2-3$), where the mean IGM transmission fraction is ~30\% \citep{Inoue14}, among LBG and LAE populations \citep{Kulas12,Yamada12,Hashimoto15}, some with particularly small peaks separation \citep{Vanzella16b}. Generally, the blue part of Lya emission has an increasing chance to be decreased ("washed out") at high-z, which means that detecting multiple peaks and measuring separation will on average become more difficult at high-z. However, for galaxies located in large ionized bubbles
-- as recently found by \citet{Castellano16}, and suggested by \citet{Matthee15}-- this may not be a problem.

Our actual sample of local LCEs allows us to calibrate diagnostics for LyC leakage from \lya\ properties. If LAEs with strong double-peaked \lya\ profiles and small separation are still detected at the highest redshift, they will be the best candidates for the sources of reionisation, and may put constraints on the patchiness of reionsation processes.

\subsection{Comparison of  \lya\ and \oiii/\oii\ diagnostics for LyC escape}

As discussed above, high \oiii/\oii\ ratios have been proposed to trace density-bounded \hii\ regions in galaxies \citep{Jaskot13,Nakajima14}. Our five LCEs were selected for their high \oiii/\oii\ ratios (and compactness), they are all leaking ionising radiation \citep{Izotov16a,Izotov16b}, and we just showed that they also have peculiar \lya\ properties (EW(\lya)$>70$\AA\ and a small peak separation), in line with our theoretical predictions \citep{Verhamme15}. So we naturally find correlations between the \oiii/\oii\ ratios of LCEs and EW(\lya) or \vsep, as illustrated in Fig.~\ref{lya_vs_O32}. But we also see in this figure that applying both criteria for LyC leakage to the comparison sample of galaxies can help selecting the best follow-up targets. 
From this exercise, three GPs -- J1219+1526 and J1424+4217 from \cite{Henry15}, and J0815+2156 from \cite{Jaskot14} -- appear as very probable LCEs, fulfilling both criteria.

\begin{figure}[htb]
\includegraphics[width=0.45\textwidth]{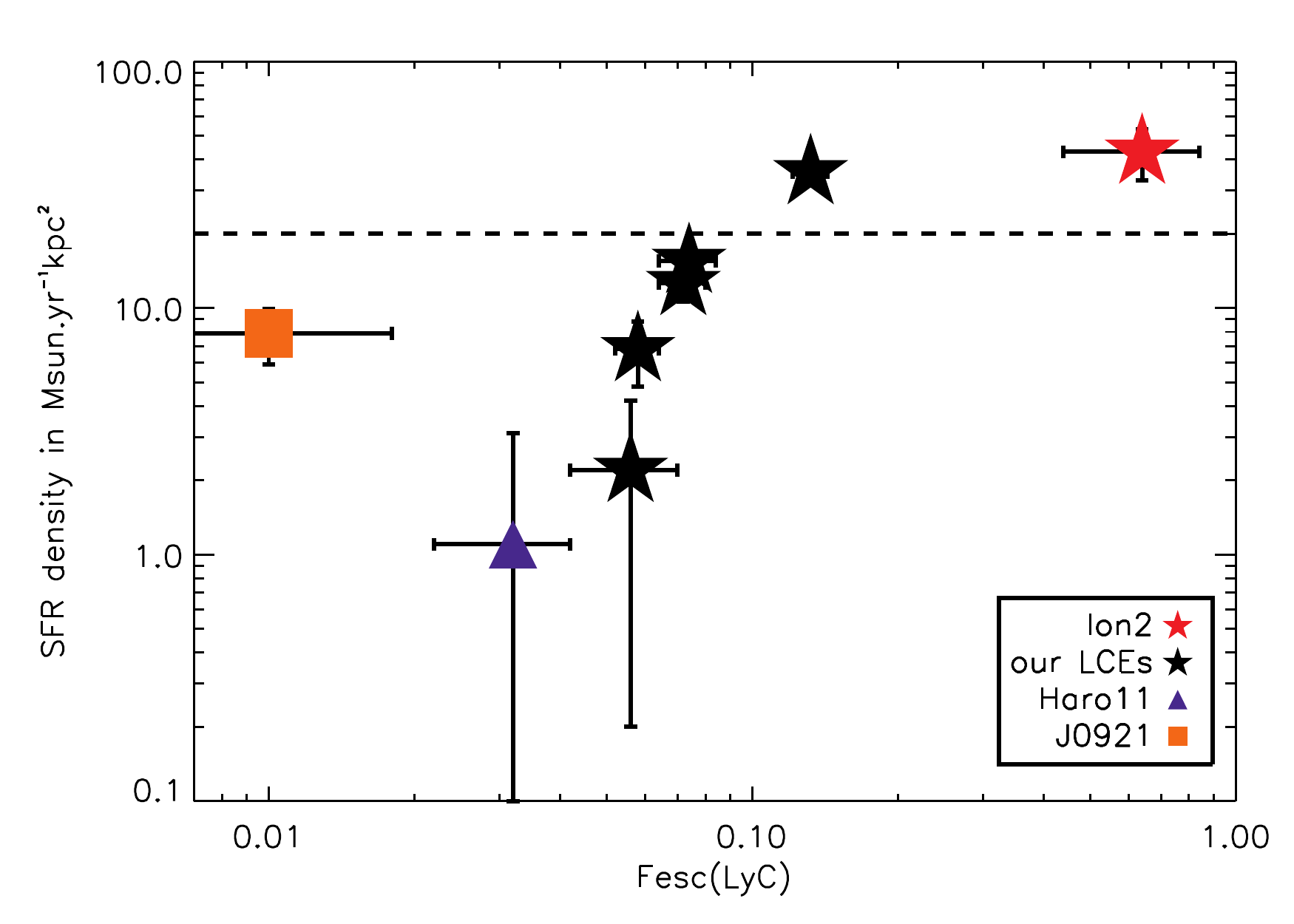} 
\caption{
Correlation between \fesclyc\ and the SFR surface density.
Same symbols as in Fig.~\protect\ref{LyaStrength}.
}
\label{sigma_vs_fescLyC}
\end{figure}

\subsection{What physical processes make these galaxies to leak ionising photons ?}

We demonstrated above that \lya\ and \oiii/\oii\ are useful diagnostics to infer LyC leakage from galaxies, or to select good LCE candidates, because both can trace low density path through the interstellar medium of galaxies along the line of sight. But these measurements do not explain the physical reason for the existence of low density paths through our galaxies. 

Our five LCEs are extremely compact, undergoing a strong star-formation episode, so they have particularly high SFR surface densities,
$\Sigma_{SFR}$. This has been proposed as a mechanism for the escape of LyC photons \citep{Heckman11, Sharma16}. Indeed, we find a strong correlation between $\Sigma_{SFR}$ and the escape fraction of ionising radiation from galaxies, \fesclyc, as shown in Fig~\ref{sigma_vs_fescLyC}. We interpret this correlation as a good indication that concentrated star-formation is necessary for LyC escape, either by producing a lot of ionising radiation from a concentration of young stars, photoionising the ISM ("density-bounded" scenario), or by mechanical feedback from these young stars (stellar winds, radiation pressure, supernova creating holes along random sight-lines), or both.

\section{Conclusions}
\label{s_conclude}

We have analyzed the properties of the \lya\ line of five low-redshift Lyman continuum leaking galaxies observed with the COS spectrograph onboard {\sl HST} that have recently been reported by \cite{Izotov16a,Izotov16b}. The $z \sim 0.3$ sources were selected for compactness and for showing a high emission line ratio \oiii/\oii\, which has previously been suggested as a possible diagnostics for Lyman continuum escape \citep{Jaskot13,Nakajima14}.
For comparison we have also included the other confirmed low-$z$ Lyman continuum emitters (LCEs), and {\sl HST}/COS observations from samples of Green Pea (GPs) galaxies and Lyman break analogs (LBAs), which have been suggested to be LCEs (see Sect.\ \ref{s_data}).
We have presented the behaviour of the \lya\ equivalent width, EW(\lya), the \lya\ escape fraction (\fesclya), and several measures of the \lya\ line profile, and discussed how these quantities depend on the Lyman continuum escape fraction, \fesclyc.

We found that strong LCEs, defined by an escape fraction \fesclyc$>5$ \%, are characterised by strong \lya\ emission, as measured by EW(\lya) and \fesclya. Furthermore, their \lya\ line profiles show a ``double-peak" structure with two emission peaks blue- and redward of the central wavelength with a small separation between the peaks ($\vsep \sim 300-400$ \kms), confirming qualitatively the predictions of \cite{Verhamme15} from radiation transfer models. More precisely, our main results can be summarised as follows:
\begin{itemize}
\item All strong LCEs show EW(\lya)$>70$ \AA\ and \fesclya $>0.2$, suggesting that these quantities could be used as efficient selection criteria to find potential sources (or analogs) of  cosmic reionisation (cf.\ Fig.~\ref{LyaStrength}).
\item The restframe EW(\lya) increases with the Lyman continuum escape fraction and reaches EW(\lya)$\sim 70-130$ \AA\ (Fig.~\ref{LyaStrength_vs_LyC} left).
\item Empirically the \lya\ escape fraction correlates with the Lyman continuum escape fraction (Fig.~\ref{LyaStrength_vs_LyC} right).
On average \fesclya/\fesclyc$\sim 2-3$, indicating that \lya\ photons escape more efficiently than the ionising photons, as expected due to radiation transfer effects (Section \ref{s_LyaLyC}).
\item The separation between the blue and red peaks decreases with increasing \fesclyc, as expected for decreasing \hi\ column densities \citep{Verhamme15}. Whereas the red peak of most LCEs is  $V_{\mathrm{peak}}^{\mathrm{red}} \sim 80-150$ \kms, the blue peak can be shifted by larger amounts ($V_{\mathrm{peak}}^{\mathrm{blue}} \sim$ -300 to -150 \kms). The shift of the blue peak is found to correlate with \fesclyc (Fig.~\ref{peaks_vs_fescLyC}).
\item No clear correlation is found for LCEs between the ratio of the blue and red equivalent widths and \fesclyc.
\item Comparison of the \lya\ line properties of GPs and LBAs with those of the LCEs shows that a subset of these
galaxies could indeed be emitters in the Lyman continuum (see Fig.~\ref{peaks_vs_EWLya}).
\item  We find a correlation between \fesclyc\ and the SFR surface density (cf.\ Fig.~\ref{sigma_vs_fescLyC}), 
indicating that the compactness of star-forming regions could play a significant role in shaping low density channels through the ISM of LCEs.
\end{itemize}

Although larger samples of LCEs are needed to establish more robust correlations between Lyman 
continuum escape and other properties, the observations reported here should provide useful guidance 
to find and select more efficiently the sources of cosmic reionisation or their analogs.


\begin{acknowledgements} 
  We thank Max Gronke for stimulating discussions and for sharing his simulation data. 
  AV thanks Mirka Dessauges for useful discussion about data reduction and spectral measurements.
  AV is supported by a Marie Heim V\"ogtlin fellowship of the Swiss National 
  Foundation. IO acknowledges the support from the grant 14-20666P of 
  Czech Science Foundation, together with the long-term institutional 
  grant RVO:67985815.
  Based on observations made with the NASA/ESA Hubble Space Telescope, 
  obtained from the data archive at the Space Telescope Science Institute.
  Support for this work was provided by NASA through grant number HST-GO-13744.001-A from the STScI.
  All of the {\sl HST} data presented in this paper were obtained from the
  Mikulski Archive for Space Telescopes (MAST). STScI is operated by
  the Association of Universities for Research in Astronomy, Inc.,
  under NASA contract NAS5-26555. Support for MAST for non-HST data is
  provided by the NASA Office of Space Science via grant NNX13AC07G and
  by other grants and contracts.
  We made use of the SAO/NASA Astrophysics Data System (ADS),
  the NASA/IPAC Extragalactic Database (NED), and the Image Reduction
  and Analysis Facility (IRAF), distributed by the
  National Optical Astronomy Observatories.
  Based on observations made with the NASA/ESA Hubble Space Telescope, 
  obtained [from the Data Archive] at the Space Telescope Science Institute, 
  which is operated by the Association of Universities for Research in Astronomy, Inc., 
  under NASA contract NAS 5-26555. These observations are associated with program 13744.
\end{acknowledgements}


\bibliographystyle{aa}
\bibliography{references}

\begin{thebibliography}{73}
\expandafter\ifx\csname natexlab\endcsname\relax\def\natexlab#1{#1}\fi

\bibitem[{{Alexandroff} {et~al.}(2015){Alexandroff}, {Heckman}, {Borthakur},
  {Overzier}, \& {Leitherer}}]{Alexandroff15}
{Alexandroff}, R.~M., {Heckman}, T.~M., {Borthakur}, S., {Overzier}, R., \&
  {Leitherer}, C. 2015, \apj, 810, 104

\bibitem[{{Atek} {et~al.}(2008){Atek}, {Kunth}, {Hayes}, {{\"O}stlin}, \&
  {Mas-Hesse}}]{Atek08}
{Atek}, H., {Kunth}, D., {Hayes}, M., {{\"O}stlin}, G., \& {Mas-Hesse}, J.~M.
  2008, \aap, 488, 491

\bibitem[{{Atek} {et~al.}(2014){Atek}, {Kunth}, {Schaerer}, {Mas-Hesse},
  {Hayes}, {{\"O}stlin}, \& {Kneib}}]{Atek14}
{Atek}, H., {Kunth}, D., {Schaerer}, D., {et~al.} 2014, \aap, 561, A89

\bibitem[{{Behrens} {et~al.}(2014){Behrens}, {Dijkstra}, \&
  {Niemeyer}}]{Behrens14}
{Behrens}, C., {Dijkstra}, M., \& {Niemeyer}, J.~C. 2014, \aap, 563, A77

\bibitem[{{Bergvall} {et~al.}(2006){Bergvall}, {Zackrisson}, {Andersson},
  {Arnberg}, {Masegosa}, \& {{\"O}stlin}}]{Bergvall06}
{Bergvall}, N., {Zackrisson}, E., {Andersson}, B.-G., {et~al.} 2006, \aap, 448,
  513

\bibitem[{{Borthakur} {et~al.}(2014){Borthakur}, {Heckman}, {Leitherer}, \&
  {Overzier}}]{Borthakur14}
{Borthakur}, S., {Heckman}, T.~M., {Leitherer}, C., \& {Overzier}, R.~A. 2014,
  Science, 346, 216

\bibitem[{{Cardamone} {et~al.}(2009){Cardamone}, {Schawinski}, {Sarzi},
  {Bamford}, {Bennert}, {Urry}, {Lintott}, {Keel}, {Parejko}, {Nichol},
  {Thomas}, {Andreescu}, {Murray}, {Raddick}, {Slosar}, {Szalay}, \&
  {Vandenberg}}]{Cardamone09}
{Cardamone}, C., {Schawinski}, K., {Sarzi}, M., {et~al.} 2009, \mnras, 399,
  1191

\bibitem[{{Castellano} {et~al.}(2016){Castellano}, {Dayal}, {Pentericci},
  {Fontana}, {Hutter}, {Brammer}, {Merlin}, {Grazian}, {Pilo}, {Amorin},
  {Cristiani}, {Dickinson}, {Ferrara}, {Gallerani}, {Giallongo}, {Giavalisco},
  {Guaita}, {Koekemoer}, {Maiolino}, {Paris}, {Santini}, {Vallini}, {Vanzella},
  \& {Wagg}}]{Castellano16}
{Castellano}, M., {Dayal}, P., {Pentericci}, L., {et~al.} 2016, \apjl, 818, L3

\bibitem[{{de Barros} {et~al.}(2016){de Barros}, {Vanzella}, {Amor{\'{\i}}n},
  {Castellano}, {Siana}, {Grazian}, {Suh}, {Balestra}, {Vignali}, {Verhamme},
  {Zamorani}, {Mignoli}, {Hasinger}, {Comastri}, {Pentericci},
  {P{\'e}rez-Montero}, {Fontana}, {Giavalisco}, \& {Gilli}}]{deBarros16}
{de Barros}, S., {Vanzella}, E., {Amor{\'{\i}}n}, R., {et~al.} 2016, \aap, 585,
  A51

\bibitem[{{Dijkstra}(2014)}]{Dijkstra14}
{Dijkstra}, M. 2014, \pasa, 31, e040

\bibitem[{{Dijkstra}(2016)}]{Dijkstra15}
{Dijkstra}, M. 2016, in Astrophysics and Space Science Library, Vol. 423,
  Astrophysics and Space Science Library, ed. A.~{Mesinger}, 145

\bibitem[{{Dijkstra} \& {Gronke}(2016)}]{Dijkstra16}
{Dijkstra}, M. \& {Gronke}, M. 2016, ArXiv e-prints

\bibitem[{{Dopita} \& {Sutherland}(2003)}]{Dopita03}
{Dopita}, M.~A. \& {Sutherland}, R.~S. 2003, {Astrophysics of the diffuse
  universe}

\bibitem[{{Duval} {et~al.}(2016){Duval}, {{\"O}stlin}, {Hayes}, {Zackrisson},
  {Verhamme}, {Orlitova}, {Adamo}, {Guaita}, {Melinder}, {Cannon}, {Laursen},
  {Rivera-Thorsen}, {Herenz}, {Gruyters}, {Mas-Hesse}, {Kunth}, {Sandberg},
  {Schaerer}, \& {M{\aa}nsson}}]{Duval16}
{Duval}, F., {{\"O}stlin}, G., {Hayes}, M., {et~al.} 2016, \aap, 587, A77

\bibitem[{{Erb} {et~al.}(2014){Erb}, {Steidel}, {Trainor}, {Bogosavljevi{\'c}},
  {Shapley}, {Nestor}, {Kulas}, {Law}, {Strom}, {Rudie}, {Reddy}, {Pettini},
  {Konidaris}, {Mace}, {Matthews}, \& {McLean}}]{Erb14}
{Erb}, D.~K., {Steidel}, C.~C., {Trainor}, R.~F., {et~al.} 2014, \apj, 795, 33

\bibitem[{{Fontanot} {et~al.}(2014){Fontanot}, {Cristiani}, {Pfrommer},
  {Cupani}, \& {Vanzella}}]{Fontanot14}
{Fontanot}, F., {Cristiani}, S., {Pfrommer}, C., {Cupani}, G., \& {Vanzella},
  E. 2014, \mnras, 438, 2097

\bibitem[{{Fontanot} {et~al.}(2012){Fontanot}, {Cristiani}, \&
  {Vanzella}}]{Fontanot12}
{Fontanot}, F., {Cristiani}, S., \& {Vanzella}, E. 2012, \mnras, 425, 1413

\bibitem[{{Fosbury} {et~al.}(2003){Fosbury}, {Villar-Mart{\'{\i}}n},
  {Humphrey}, {Lombardi}, {Rosati}, {Stern}, {Hook}, {Holden}, {Stanford},
  {Squires}, {Rauch}, \& {Sargent}}]{Fosbury03}
{Fosbury}, R.~A.~E., {Villar-Mart{\'{\i}}n}, M., {Humphrey}, A., {et~al.} 2003,
  \apj, 596, 797

\bibitem[{{Giallongo} {et~al.}(2015){Giallongo}, {Grazian}, {Fiore}, {Fontana},
  {Pentericci}, {Vanzella}, {Dickinson}, {Kocevski}, {Castellano}, {Cristiani},
  {Ferguson}, {Finkelstein}, {Grogin}, {Hathi}, {Koekemoer}, {Newman}, \&
  {Salvato}}]{Giallongo15}
{Giallongo}, E., {Grazian}, A., {Fiore}, F., {et~al.} 2015, \aap, 578, A83

\bibitem[{{Gronke} \& {Dijkstra}(2016)}]{Gronke16}
{Gronke}, M. \& {Dijkstra}, M. 2016, \apj, 826, 14

\bibitem[{{Haiman}(2002)}]{Haiman02}
{Haiman}, Z. 2002, \apjl, 576, L1

\bibitem[{{Hashimoto} {et~al.}(2015){Hashimoto}, {Verhamme}, {Ouchi},
  {Shimasaku}, {Schaerer}, {Nakajima}, {Shibuya}, {Rauch}, {Ono}, \&
  {Goto}}]{Hashimoto15}
{Hashimoto}, T., {Verhamme}, A., {Ouchi}, M., {et~al.} 2015, \apj, 812, 157

\bibitem[{{Hayes} {et~al.}(2014){Hayes}, {{\"O}stlin}, {Duval}, {Sandberg},
  {Guaita}, {Melinder}, {Adamo}, {Schaerer}, {Verhamme}, {Orlitov{\'a}},
  {Mas-Hesse}, {Cannon}, {Atek}, {Kunth}, {Laursen}, {Ot{\'{\i}}-Floranes},
  {Pardy}, {Rivera-Thorsen}, \& {Herenz}}]{Hayes14}
{Hayes}, M., {{\"O}stlin}, G., {Duval}, F., {et~al.} 2014, \apj, 782, 6

\bibitem[{{Hayes} {et~al.}(2013){Hayes}, {{\"O}stlin}, {Schaerer}, {Verhamme},
  {Mas-Hesse}, {Adamo}, {Atek}, {Cannon}, {Duval}, {Guaita}, {Herenz}, {Kunth},
  {Laursen}, {Melinder}, {Orlitov{\'a}}, {Ot{\'{\i}}-Floranes}, \&
  {Sandberg}}]{Hayes13}
{Hayes}, M., {{\"O}stlin}, G., {Schaerer}, D., {et~al.} 2013, \apjl, 765, L27

\bibitem[{{Hayes} {et~al.}(2011){Hayes}, {Schaerer}, {{\"O}stlin}, {Mas-Hesse},
  {Atek}, \& {Kunth}}]{Hayes11}
{Hayes}, M., {Schaerer}, D., {{\"O}stlin}, G., {et~al.} 2011, \apj, 730, 8

\bibitem[{{Heckman} {et~al.}(2011){Heckman}, {Borthakur}, {Overzier},
  {Kauffmann}, {Basu-Zych}, {Leitherer}, {Sembach}, {Martin}, {Rich},
  {Schiminovich}, \& {Seibert}}]{Heckman11}
{Heckman}, T.~M., {Borthakur}, S., {Overzier}, R., {et~al.} 2011, \apj, 730, 5

\bibitem[{{Heckman} {et~al.}(2005){Heckman}, {Hoopes}, {Seibert}, {Martin},
  {Salim}, {Rich}, {Kauffmann}, {Charlot}, {Barlow}, {Bianchi}, {Byun},
  {Donas}, {Forster}, {Friedman}, {Jelinsky}, {Lee}, {Madore}, {Malina},
  {Milliard}, {Morrissey}, {Neff}, {Schiminovich}, {Siegmund}, {Small},
  {Szalay}, {Welsh}, \& {Wyder}}]{Heckman05}
{Heckman}, T.~M., {Hoopes}, C.~G., {Seibert}, M., {et~al.} 2005, \apjl, 619,
  L35

\bibitem[{{Henry} {et~al.}(2015){Henry}, {Scarlata}, {Martin}, \&
  {Erb}}]{Henry15}
{Henry}, A., {Scarlata}, C., {Martin}, C.~L., \& {Erb}, D. 2015, \apj, 809, 19

\bibitem[{{Inoue} {et~al.}(2014){Inoue}, {Shimizu}, {Iwata}, \&
  {Tanaka}}]{Inoue14}
{Inoue}, A.~K., {Shimizu}, I., {Iwata}, I., \& {Tanaka}, M. 2014, \mnras, 442,
  1805

\bibitem[{{Izotov} {et~al.}(2011){Izotov}, {Guseva}, \& {Thuan}}]{Izotov11}
{Izotov}, Y.~I., {Guseva}, N.~G., \& {Thuan}, T.~X. 2011, \apj, 728, 161

\bibitem[{{Izotov} {et~al.}(2016{\natexlab{a}}){Izotov}, {Orlitov{\'a}},
  {Schaerer}, {Thuan}, {Verhamme}, {Guseva}, \& {Worseck}}]{Izotov16a}
{Izotov}, Y.~I., {Orlitov{\'a}}, I., {Schaerer}, D., {et~al.}
  2016{\natexlab{a}}, \nat, 529, 178

\bibitem[{{Izotov} {et~al.}(2016{\natexlab{b}}){Izotov}, {Schaerer}, {Thuan},
  {Worseck}, {Guseva}, {Orlitov{\'a}}, \& {Verhamme}}]{Izotov16b}
{Izotov}, Y.~I., {Schaerer}, D., {Thuan}, T.~X., {et~al.} 2016{\natexlab{b}},
  \mnras, 461, 3683

\bibitem[{{Jaskot} \& {Oey}(2013)}]{Jaskot13}
{Jaskot}, A.~E. \& {Oey}, M.~S. 2013, \apj, 766, 91

\bibitem[{{Jaskot} \& {Oey}(2014)}]{Jaskot14}
{Jaskot}, A.~E. \& {Oey}, M.~S. 2014, \apjl, 791, L19

\bibitem[{{Kulas} {et~al.}(2012){Kulas}, {Shapley}, {Kollmeier}, {Zheng},
  {Steidel}, \& {Hainline}}]{Kulas12}
{Kulas}, K.~R., {Shapley}, A.~E., {Kollmeier}, J.~A., {et~al.} 2012, \apj, 745,
  33

\bibitem[{{Laursen} {et~al.}(2011){Laursen}, {Sommer-Larsen}, \&
  {Razoumov}}]{Laursen11}
{Laursen}, P., {Sommer-Larsen}, J., \& {Razoumov}, A.~O. 2011, \apj, 728, 52

\bibitem[{{Leitet} {et~al.}(2013){Leitet}, {Bergvall}, {Hayes}, {Linn{\'e}}, \&
  {Zackrisson}}]{Leitet13}
{Leitet}, E., {Bergvall}, N., {Hayes}, M., {Linn{\'e}}, S., \& {Zackrisson}, E.
  2013, \aap, 553, A106

\bibitem[{{Leitet} {et~al.}(2011){Leitet}, {Bergvall}, {Piskunov}, \&
  {Andersson}}]{Leitet11}
{Leitet}, E., {Bergvall}, N., {Piskunov}, N., \& {Andersson}, B.-G. 2011, \aap,
  532, A107

\bibitem[{{Leitherer} {et~al.}(2016){Leitherer}, {Hernandez}, {Lee}, \&
  {Oey}}]{Leitherer16}
{Leitherer}, C., {Hernandez}, S., {Lee}, J.~C., \& {Oey}, M.~S. 2016, ArXiv
  e-prints

\bibitem[{{Madau}(1995)}]{Madau95}
{Madau}, P. 1995, \apj, 441, 18

\bibitem[{{Madau} \& {Haardt}(2015)}]{Madau15}
{Madau}, P. \& {Haardt}, F. 2015, \apjl, 813, L8

\bibitem[{{Matthee} {et~al.}(2016{\natexlab{a}}){Matthee}, {Sobral}, {Best},
  {Khostovan}, {Oteo}, {Bouwens}, \& {R{\"o}ttgering}}]{Matthee16b}
{Matthee}, J., {Sobral}, D., {Best}, P., {et~al.} 2016{\natexlab{a}}, ArXiv
  e-prints

\bibitem[{{Matthee} {et~al.}(2016{\natexlab{b}}){Matthee}, {Sobral}, {Oteo},
  {Best}, {Smail}, {R{\"o}ttgering}, \& {Paulino-Afonso}}]{Matthee16a}
{Matthee}, J., {Sobral}, D., {Oteo}, I., {et~al.} 2016{\natexlab{b}}, \mnras,
  458, 449

\bibitem[{{Matthee} {et~al.}(2015){Matthee}, {Sobral}, {Santos},
  {R{\"o}ttgering}, {Darvish}, \& {Mobasher}}]{Matthee15}
{Matthee}, J., {Sobral}, D., {Santos}, S., {et~al.} 2015, \mnras, 451, 400

\bibitem[{{Micheva} {et~al.}(2016){Micheva}, {Iwata}, \& {Inoue}}]{Micheva16}
{Micheva}, G., {Iwata}, I., \& {Inoue}, A.~K. 2016, ArXiv e-prints

\bibitem[{{Nakajima} \& {Ouchi}(2014)}]{Nakajima14}
{Nakajima}, K. \& {Ouchi}, M. 2014, \mnras, 442, 900

\bibitem[{{Ocvirk} {et~al.}(2015){Ocvirk}, {Gillet}, {Shapiro}, {Aubert},
  {Iliev}, {Teyssier}, {Yepes}, {Choi}, {Sullivan}, {Knebe}, {Gottloeber},
  {D'Aloisio}, {Park}, {Hoffman}, \& {Stranex}}]{Ocvirk16}
{Ocvirk}, P., {Gillet}, N., {Shapiro}, P.~R., {et~al.} 2015, ArXiv e-prints

\bibitem[{{{\"O}stlin} {et~al.}(2014){{\"O}stlin}, {Hayes}, {Duval},
  {Sandberg}, {Rivera-Thorsen}, {Marquart}, {Orlitov{\'a}}, {Adamo},
  {Melinder}, {Guaita}, {Atek}, {Cannon}, {Gruyters}, {Herenz}, {Kunth},
  {Laursen}, {Mas-Hesse}, {Micheva}, {Ot{\'{\i}}-Floranes}, {Pardy}, {Roth},
  {Schaerer}, \& {Verhamme}}]{Ostlin14}
{{\"O}stlin}, G., {Hayes}, M., {Duval}, F., {et~al.} 2014, \apj, 797, 11

\bibitem[{{{\"O}stlin} {et~al.}(2009){{\"O}stlin}, {Hayes}, {Kunth},
  {Mas-Hesse}, {Leitherer}, {Petrosian}, \& {Atek}}]{Ostlin09}
{{\"O}stlin}, G., {Hayes}, M., {Kunth}, D., {et~al.} 2009, \aj, 138, 923

\bibitem[{{Overzier} {et~al.}(2009){Overzier}, {Heckman}, {Tremonti}, {Armus},
  {Basu-Zych}, {Gon{\c c}alves}, {Rich}, {Martin}, {Ptak}, {Schiminovich},
  {Ford}, {Madore}, \& {Seibert}}]{Overzier09}
{Overzier}, R.~A., {Heckman}, T.~M., {Tremonti}, C., {et~al.} 2009, \apj, 706,
  203

\bibitem[{{Pe{\~n}a-Guerrero} \& {Leitherer}(2013)}]{Pena13}
{Pe{\~n}a-Guerrero}, M.~A. \& {Leitherer}, C. 2013, \aj, 146, 158

\bibitem[{{Rivera-Thorsen} {et~al.}(2015){Rivera-Thorsen}, {Hayes},
  {{\"O}stlin}, {Duval}, {Orlitov{\'a}}, {Verhamme}, {Mas-Hesse}, {Schaerer},
  {Cannon}, {Ot{\'{\i}}-Floranes}, {Sandberg}, {Guaita}, {Adamo}, {Atek},
  {Herenz}, {Kunth}, {Laursen}, \& {Melinder}}]{Rivera15}
{Rivera-Thorsen}, T.~E., {Hayes}, M., {{\"O}stlin}, G., {et~al.} 2015, \apj,
  805, 14

\bibitem[{{Santos}(2004)}]{Santos04}
{Santos}, M.~R. 2004, \mnras, 349, 1137

\bibitem[{{Schaerer}(2003)}]{Schaerer03}
{Schaerer}, D. 2003, \aap, 397, 527

\bibitem[{{Schaerer} \& {Verhamme}(2008)}]{Schaerer08}
{Schaerer}, D. \& {Verhamme}, A. 2008, \aap, 480, 369

\bibitem[{{Schenker} {et~al.}(2014){Schenker}, {Ellis}, {Konidaris}, \&
  {Stark}}]{Schenker14}
{Schenker}, M.~A., {Ellis}, R.~S., {Konidaris}, N.~P., \& {Stark}, D.~P. 2014,
  \apj, 795, 20

\bibitem[{{Shapley} {et~al.}(2003){Shapley}, {Steidel}, {Pettini}, \&
  {Adelberger}}]{Shapley03}
{Shapley}, A.~E., {Steidel}, C.~C., {Pettini}, M., \& {Adelberger}, K.~L. 2003,
  \apj, 588, 65

\bibitem[{{Shapley} {et~al.}(2016){Shapley}, {Steidel}, {Strom},
  {Bogosavljevi{\'c}}, {Reddy}, {Siana}, {Mostardi}, \& {Rudie}}]{Shapley16}
{Shapley}, A.~E., {Steidel}, C.~C., {Strom}, A.~L., {et~al.} 2016, ArXiv
  e-prints

\bibitem[{{Sharma} {et~al.}(2016){Sharma}, {Theuns}, {Frenk}, {Bower}, {Crain},
  {Schaller}, \& {Schaye}}]{Sharma16}
{Sharma}, M., {Theuns}, T., {Frenk}, C., {et~al.} 2016, \mnras, 458, L94

\bibitem[{{Stark} {et~al.}(2011){Stark}, {Ellis}, \& {Ouchi}}]{Stark11}
{Stark}, D.~P., {Ellis}, R.~S., \& {Ouchi}, M. 2011, \apjl, 728, L2

\bibitem[{{Steidel} {et~al.}(2011){Steidel}, {Bogosavljevi{\'c}}, {Shapley},
  {Kollmeier}, {Reddy}, {Erb}, \& {Pettini}}]{Steidel11}
{Steidel}, C.~C., {Bogosavljevi{\'c}}, M., {Shapley}, A.~E., {et~al.} 2011,
  \apj, 736, 160

\bibitem[{{Syphers} {et~al.}(2012){Syphers}, {Anderson}, {Zheng}, {Meiksin},
  {Schneider}, \& {York}}]{Syphers12}
{Syphers}, D., {Anderson}, S.~F., {Zheng}, W., {et~al.} 2012, \aj, 143, 100

\bibitem[{{Vanzella} {et~al.}(2015){Vanzella}, {de Barros}, {Castellano},
  {Grazian}, {Inoue}, {Schaerer}, {Guaita}, {Zamorani}, {Giavalisco}, {Siana},
  {Pentericci}, {Giallongo}, {Fontana}, \& {Vignali}}]{Vanzella15}
{Vanzella}, E., {de Barros}, S., {Castellano}, M., {et~al.} 2015, \aap, 576,
  A116

\bibitem[{{Vanzella} {et~al.}(2016{\natexlab{a}}){Vanzella}, {De Barros},
  {Cupani}, {Karman}, {Gronke}, {Balestra}, {Coe}, {Mignoli}, {Brusa},
  {Calura}, {Caminha}, {Caputi}, {Castellano}, {Christensen}, {Comastri},
  {Cristiani}, {Dijkstra}, {Fontana}, {Giallongo}, {Giavalisco}, {Gilli},
  {Grazian}, {Grillo}, {Koekemoer}, {Meneghetti}, {Nonino}, {Pentericci},
  {Rosati}, {Schaerer}, {Verhamme}, {Vignali}, \& {Zamorani}}]{Vanzella16b}
{Vanzella}, E., {De Barros}, S., {Cupani}, G., {et~al.} 2016{\natexlab{a}},
  \apjl, 821, L27

\bibitem[{{Vanzella} {et~al.}(2016{\natexlab{b}}){Vanzella}, {de Barros},
  {Vasei}, {Alavi}, {Giavalisco}, {Siana}, {Grazian}, {Hasinger}, {Suh},
  {Cappelluti}, {Vito}, {Amorin}, {Balestra}, {Brusa}, {Calura}, {Castellano},
  {Comastri}, {Fontana}, {Gilli}, {Mignoli}, {Pentericci}, {Vignali}, \&
  {Zamorani}}]{Vanzella16}
{Vanzella}, E., {de Barros}, S., {Vasei}, K., {et~al.} 2016{\natexlab{b}},
  \apj, 825, 41

\bibitem[{{Vanzella} {et~al.}(2010){Vanzella}, {Giavalisco}, {Inoue}, {Nonino},
  {Fontanot}, {Cristiani}, {Grazian}, {Dickinson}, {Stern}, {Tozzi},
  {Giallongo}, {Ferguson}, {Spinrad}, {Boutsia}, {Fontana}, {Rosati}, \&
  {Pentericci}}]{Vanzella10}
{Vanzella}, E., {Giavalisco}, M., {Inoue}, A.~K., {et~al.} 2010, \apj, 725,
  1011

\bibitem[{{Vasei} {et~al.}(2016){Vasei}, {Siana}, {Shapley}, {Quider}, {Alavi},
  {Rafelski}, {Steidel}, {Pettini}, \& {Lewis}}]{Vasei16}
{Vasei}, K., {Siana}, B., {Shapley}, A.~E., {et~al.} 2016, ArXiv e-prints

\bibitem[{{Verhamme} {et~al.}(2015){Verhamme}, {Orlitov{\'a}}, {Schaerer}, \&
  {Hayes}}]{Verhamme15}
{Verhamme}, A., {Orlitov{\'a}}, I., {Schaerer}, D., \& {Hayes}, M. 2015, \aap,
  578, A7

\bibitem[{{Wofford} {et~al.}(2013){Wofford}, {Leitherer}, \&
  {Salzer}}]{Wofford13}
{Wofford}, A., {Leitherer}, C., \& {Salzer}, J. 2013, \apj, 765, 118

\bibitem[{{Worseck} {et~al.}(2011){Worseck}, {Prochaska}, {McQuinn},
  {Dall'Aglio}, {Fechner}, {Hennawi}, {Reimers}, {Richter}, \&
  {Wisotzki}}]{Worseck11}
{Worseck}, G., {Prochaska}, J.~X., {McQuinn}, M., {et~al.} 2011, \apjl, 733,
  L24

\bibitem[{{Yamada} {et~al.}(2012){Yamada}, {Matsuda}, {Kousai}, {Hayashino},
  {Morimoto}, \& {Umemura}}]{Yamada12}
{Yamada}, T., {Matsuda}, Y., {Kousai}, K., {et~al.} 2012, \apj, 751, 29

\bibitem[{{Yang} {et~al.}(2016){Yang}, {Malhotra}, {Gronke}, {Rhoads},
  {Dijkstra}, {Jaskot}, {Zheng}, \& {Wang}}]{Yang16}
{Yang}, H., {Malhotra}, S., {Gronke}, M., {et~al.} 2016, \apj, 820, 130

\bibitem[{{Zackrisson} {et~al.}(2013){Zackrisson}, {Inoue}, \&
  {Jensen}}]{Zackrisson13}
{Zackrisson}, E., {Inoue}, A.~K., \& {Jensen}, H. 2013, \apj, 777, 39

\end{thebibliography}

\end{document}